%% file: main.tex
\documentclass[a4paper,fleqn]{cas-sc}

\usepackage[numbers]{natbib}
\usepackage{graphicx}
\usepackage{amsmath,amssymb,amsfonts}
\usepackage[ruled,linesnumbered,noend]{algorithm2e}
\usepackage{textcomp}
\usepackage{xcolor}
\usepackage{comment}
\usepackage{breakurl}
\usepackage{xspace}
\usepackage{subfigure}
\usepackage{pifont}
\usepackage{multirow}
\usepackage{amsfonts}
\usepackage{listings}
\usepackage{subcaption}
\usepackage{color}
\usepackage{url}
\usepackage{booktabs}
\usepackage{colortbl}
\usepackage{tcolorbox}
\usepackage{hyperref}
\usepackage{fvextra}
\usepackage{enumitem}
\usepackage{listings}
\usepackage{cases}
\usepackage{textcomp}

\newcommand{\yg}[1]{\textcolor{black}{#1}}

\definecolor{light-gray}{gray}{0.95}

\newcommand{\tool}{\textsf{Flab-Pruner}}

\definecolor{dkgreen}{rgb}{0,0.6,0}
\definecolor{gray}{rgb}{0.5,0.5,0.5}
\definecolor{mauve}{rgb}{0.58,0,0.82}

\newcommand{\best}{gray!30}

\def\tsc#1{\csdef{#1}{\textsc{\lowercase{#1}}\xspace}}
\tsc{WGM}
\tsc{QE}
\tsc{EP}
\tsc{PMS}
\tsc{BEC}
\tsc{DE}

\begin{document}
\let\WriteBookmarks\relax
\def\floatpagepagefraction{1}
\def\textpagefraction{.001}

\shorttitle{Less is More: Towards Green Code Large Language Models via Unified Structural Pruning}


\title [mode = title]{Less is More: Towards Green Code Large Language Models via Unified Structural Pruning}



%

\author[1, 2]{Guang Yang}



\ead{yang.guang@nuaa.edu.cn}

\ead[url]{https://ntdxyg.github.io/}

\credit{Data curation, Methodology, Software, Writing - Original draft preparation}

\author[1]{Yu Zhou}
\cormark[1]
\ead{zhouyu@nuaa.edu.cn}
\ead[url]{https://csyuzhou.github.io/}
\credit{Conceptualization, Methodology, Writing - Review \& Editing}

\author[1]{Xiangyu Zhang}
\ead{zhangx1angyu@nuaa.edu.cn}
\credit{Data curation, Writing - Review \& Editing}

\author[1]{Wei Cheng}
\ead{chengweii@nuaa.edu.cn}
\credit{Validation, Writing - Review \& Editing}

\author[3]{Ke Liu}
\ead{liuke23@nudt.edu.cn}
\credit{Validation, Writing - Review \& Editing}

\author[4]{Xiang Chen}
\ead{xchencs@ntu.edu.cn}
\ead[url]{https://smartse.github.io/}
\credit{Validation, Writing - Review \& Editing}

\author[5]{Terry Yue Zhuo}
\ead{terry.zhuo@monash.edu}
\ead[url]{https://terryyz.github.io/}
\credit{Methodology, Writing - Review \& Editing}

\author[6]{Taolue Chen}
\cormark[1]
\ead{taolue.chen@gmail.com}
\ead[url]{https://chentaolue.github.io/}
\credit{Conceptualization, Methodology, Writing - Review \& Editing, Validation}

\affiliation[1]{organization={Nanjing University of Aeronautics and Astronautics},
city={Nanjing},
country={China}}

\affiliation[2]{organization={Singapore Management University},
city={Singapore},
country={Singapore}}

\affiliation[3]{organization={National University of Defense Technology},
city={Changsha},
country={China}}

\affiliation[4]{organization={Nantong University},
city={Nantong},
country={China}}

\affiliation[5]{organization={Monash University and CSIRO's Data61},
city={Melbourne},
country={Australia}}

\affiliation[6]{organization={Birkbeck, University of London},
city={London},
country={UK}}

\cortext[cor1]{Corresponding author}

\begin{abstract}
The extensive application of Large Language Models (LLMs) in generative coding tasks has raised concerns due to their high computational demands and energy consumption. 
Unlike previous structural pruning methods designed for classification models that deal with low-dimensional classification logits, generative Code LLMs produce high-dimensional token logit sequences, making traditional pruning objectives inherently limited. 
Moreover, existing single-component pruning approaches further constrain the effectiveness when applied to generative Code LLMs. 
In response, we propose {\tool}, an innovative unified structural pruning method that combines vocabulary, layer, and Feed-Forward Network (FFN) pruning. 
This approach effectively reduces model parameters while maintaining performance. 
Additionally, we introduce a customized code instruction data strategy for coding tasks to enhance the performance recovery efficiency of the pruned model. 
Through extensive evaluations on three state-of-the-art Code LLMs across multiple generative coding tasks, the results demonstrate that {\tool} retains 97\% of the original performance after pruning 22\% of the parameters and achieves the same or even better performance after post-training. 
The pruned models exhibit significant improvements in storage, GPU usage, computational efficiency, and environmental impact, while maintaining well robustness. Our research provides a sustainable solution for green software engineering and promotes the efficient deployment of LLMs in real-world generative coding intelligence applications.
\end{abstract}



\begin{keywords}
Large Language Models \sep 
Code Intelligence \sep 
Structural Pruning \sep 
Post-Training \sep 
Code Instruction Tuning
\end{keywords}

\maketitle


\input{sections/1intro}

\input{sections/2background}
\input{sections/3method}
\input{sections/4setup.tex}
\input{sections/5results.tex}

\input{sections/6.threat.tex}
\input{sections/7.relate.tex}
\input{sections/8.conclusion.tex}

\section*{Acknowledgement}

This work is partially supported by the National Natural Science Foundation of China (NSFC, No.\ 61972197 and No.\ 62372232), the Natural Science Foundation of Jiangsu Province (No.\ BK20201292), the Collaborative Innovation Center of Novel Software Technology and Industrialization, and the Short-term Visiting Program of Nanjing University of Aeronautics and Astronautics for Ph.D. Students Abroad (No.\ 240501DF16).
T.\ Chen is partially supported by an oversea grant from the State Key Laboratory of Novel Software Technology, Nanjing University (KFKT2022A03).

\section*{Declaration of Competing Interests}
The authors declare that they have no known competing financial interests or personal relationships that could have appeared to influence the work reported in this paper.

\printcredits

\bibliographystyle{cas-model2-names}

\bibliography{cas-refs}

\end{document}

%% file: sections/1intro.tex
\section{Introduction}
\label{sec:intro}

Large Language Models (LLMs) have demonstrated outstanding performance and been deployed across numerous domains~\cite{yuan2023revisiting, huang2023chatgpt, yao2024survey}. 
Software engineering is no exception~\cite{hou2023large, kirova2024software, sallou2024breaking, fan2023large}, with LLMs excelling in tasks like code generation~\cite{gu2023llm}, summarization~\cite{nam2024using}, and vulnerability detection~\cite{lu2024grace}. 
However, the substantial scale and intensive computational requirements of these models pose challenges, particularly in resource-constrained environments~\cite{chen2020deep, hort2023exploratory, liu2013gearing}. 
Moreover, the energy consumption associated with training and inference leads to high carbon emissions, raising concerns about environmental sustainability~\cite{shi2024greening, shi2024efficient}.

To enhance energy efficiency and sustainability, green software engineering has spurred exploration of model pruning, quantization, and knowledge distillation~\cite{liu2023tcra, zhu2023survey, wei2023towards, wang2024model}, aimed at reducing the computational and environmental impact of LLMs. 
Quantization speeds up inference by converting high-precision weights to lower precision. 
Knowledge distillation transfers knowledge from larger to smaller models, requiring additional computational resources. 
Among these, model pruning has emerged as a promising strategy, with two main approaches: unstructured pruning that merely zeros out specific weights while maintaining the original parameter count, and structured pruning~\cite{wang2020structured} that effectively reduces model size by removing entire structural components (e.g., neurons, layers) while preserving model integrity and functionality.
For instance, unstructured pruning, such as SparseGPT~\cite{frantar2023sparsegpt}, targets individual weights, achieving sparsity but not significantly reducing hardware demands, thus limiting their use in constrained settings. 
In contrast, structured pruning, such as ShortGPT~\cite{men2024shortgpt}, reduces parameter counts and hardware dependencies, enhancing operational efficiency while maintaining model integrity. 

Although these pruning methods can be theoretically applied to code generation models, our extensive experiments (cf.\ RQ1 in Section~\ref{subsect:rq1}) reveal their significant limitations when applied to generative coding tasks. 
For example, when applying ShortGPT~\cite{men2024shortgpt} to the LLAMA model, we observe a complete performance collapse on the HumanEval code generation benchmark. 
Through careful analysis, we identify two fundamental limitations in existing approaches:
\begin{enumerate}
    \item \textbf{Misaligned Pruning Objective:} Current pruning methods~\cite{men2024shortgpt, yang2024laco, kim2024shortened} primarily focus on layer-wise similarity metrics (e.g., angle distance, cosine similarity, and Taylor score between layers), which aim to preserve the model's general language modeling capabilities while overlooking the specific requirements of downstream tasks. 
    
    \item \textbf{Limited Pruning Scope:} Existing approaches typically adopt single-component pruning strategies (e.g., solely focusing on layer redundancy reduction), failing to leverage the potential synergies that could be achieved through an integrated, multi-granular pruning approach across different model components.
    
    \item \textbf{Insufficient Code-specific Post-tuning Method:} Existing pruning methods rely on generic supervised fine-tuning on downstream datasets for performance recovery, without considering domain-specific adaptations for code-related tasks. This generic approach fails to leverage the unique characteristics and requirements of code generation tasks, potentially limiting the effectiveness of the post-pruning recovery process.

\end{enumerate}
We provide a detailed analysis of these limitations and their implications in Section~\ref{subsect:rq1}.

Beyond performance considerations, the reliability and robustness of pruned models in coding tasks raise critical concerns for real-world deployment. 
These concerns encompass several key aspects: How well do pruned models maintain their robustness against adversarial inputs? 
Despite the crucial nature of these questions for practical applications, existing research has largely focused on performance metrics while leaving these reliability aspects unexplored.
We present a comprehensive analysis of these critical concerns in Section~\ref{subsect:rq3}.

\smallskip
\noindent\textbf{Method.}
To minimize model parameters while ensuring that pruned models maintain high standards in performance and robustness,
First, we define the pruning objectives: since the model generates code through probability distributions over the vocabulary, we use KL divergence to ensure that the pruned model maintains similar token generation probabilities as the original model, directly optimizing for code generation behavior.
Then we introduce {\tool}, a unified structural pruning method designed for the combination of three components, i.e., \underline{\textbf{F}}FN Pruning, \underline{\textbf{La}}yer Pruning and Voca\underline{\textbf{b}}ulary Pruning. 
In particular, the vocabulary pruning component reduces the model's embedding size by eliminating tokens that are absent in the given programming corpus.
The FFN pruning component targets specific neurons within the FFN block, reducing the model's size by eliminating certain neurons.
The layer pruning component reduces the number of layers in the model by assessing the redundancy between layers.
To consider the consistency between pruning objectives and downstream task performance, the above pruning components are all designed to minimize the KL divergence between the pruned model and the original model.

Additionally, we introduce a customized code instruction tuning strategy specifically designed for generative coding tasks. 
In contrast to directly train the pruned model on the original dataset, we purposefully replace the code generated by the original model into the training dataset by evaluating the performance of the original and pruned models on the training set.
Compared to using the original dataset for performance recovery, our approach enhances efficiency by achieving better performance recovery.


\smallskip
\noindent\textbf{Evaluation.} 
We undertake a comprehensive evaluation of {\tool} spanning three widely studied code intelligence tasks, i.e., code generation, CoT generation and code output prediction. 
The goal of code generation is to convert requirement in natural language into code, bridging the gap between description and execution. 
CoT generation is about creating thought sequences from prompts, showcasing advanced reasoning. 
Code output prediction assesses the model's understanding of code and its ability to predict outcomes.
By considering these diverse tasks, our evaluation aims to provide a holistic evaluation of {\tool}'s capabilities in various code intelligence domains.
Our evaluation targets CodeQwen-1.5\footnote{\url{https://huggingface.co/Qwen/CodeQwen1.5-7B-Chat}} and its two variants (NxCode\footnote{\url{https://huggingface.co/NTQAI/Nxcode-CQ-7B-orpo}} and CodeSlerp\footnote{\url{https://huggingface.co/dohrisalim/Code-7B-slerp}}), which are top-performing 7B-10B models on the BigCode benchmark\footnote{\url{https://huggingface.co/spaces/bigcode/bigcode-models-leaderboard}}.

Besides performance, we conduct efficiency and robustness analyses to provide a comprehensive evaluation. An ablation study is also performed to evaluate the contribution of each component to our pruning methodology, highlighting their synergistic effects.

\smallskip
\noindent\textbf{Findings.}
By applying {\tool}, we demonstrate the feasibility of pruning Code LLMs while preserving their core capabilities. Our empirical findings indicate that by pruning about 22\% of the parameters, the pruned model retains around 97\% of the original model's performance.
After performance recovery, the pruned model achieves comparable or even superior performance to the original model. 
Significant improvements are observed across various efficiency metrics, including reduced storage requirements, optimized GPU utilization, decreased FLOPs, lower CO$_2$ emissions, and increased tokens processed per second. 
Moreover, our structured pruning technique is orthogonal to existing quantization methods, enabling further efficiency gains through their combination.
Furthermore, we demonstrate that the pruned model maintains its robustness across different perturbation scenarios, affirming the effectiveness of our pruning methodology.


The main contributions can be summarized as follows.

\begin{itemize}
\item  We introduce {\tool}, a novel unified structural pruning approach that integrates FFN pruning, layer pruning, and vocabulary pruning to effectively minimize model size while maintaining performance for generative coding tasks.

\item We propose a customized training data instruction strategy specifically designed for coding tasks, ensuring that pruned models continue to perform effectively.

\item We conduct an extensive evaluation of {\tool} across multiple generative coding tasks in terms of the performance, efficiency, and robustness, demonstrating the practical applicability of our approach in real-world software engineering scenarios.

\end{itemize}

To our best knowledge, this is one of the first attempts to comprehensively assess the impact of pruning techniques on the performance of LLMs in generative coding tasks.
To facilitate the replication, the 
pruned models\footnote{\url{https://www.modelscope.cn/profile/FlabPruner}} and datasets\footnote{\url{https://huggingface.co/datasets/Flab-Pruner/CodeHarmony}} are all made publicly available.

\smallskip
\noindent\textbf{Organization.} The rest of the paper is structured as follows. 
Section~\ref{sec:background} provides the background. 
In Section~\ref{sec:method}, we detail our proposed method for vocabulary pruning, layer pruning and FFN pruning. Section~\ref{sec:setup} describes the experiment settings. 
Section~\ref{sec:result} presents the experimental results. 
Section~\ref{sec:threats} analyzes threats to validity and  Section~\ref{sec:relate} discusses relevant studies.
Finally, we conclude our paper and present directions of future work in Section~\ref{sec:conclusion}.

%% file: sections/2background.tex
\section{Background}
\label{sec:background}
   
\subsection{Transformer}
A standard LLM, such as the Transformer~\cite{vaswani2017attention}, processes input sequences through several key components, including an embedding layer, multiple self-attention layers, feed-forward networks, and an output layer.

Given an input sequence of token indices \( X = (x_1, x_2, \ldots, x_n) \), the LLM first maps these discrete indices into continuous vectors using an embedding layer:

\begin{equation}
    E = (e_1, e_2, \ldots, e_n) \quad \text{, where} \quad e_i = W_e[x_i]
\end{equation}

Here, \( W_e \in \mathbb{R}^{|V| \times d} \) is the embedding matrix, \( |V| \) is the vocabulary size, and \( d \) is the dimension of the embeddings. Each token \( x_i \) is mapped to its corresponding embedding vector \( e_i \).
The embedded sequence \( E \) is then processed through a series of self-attention layers and feed-forward networks. For each layer \( l \), the transformation from input \( H^{(l-1)} \) to output \( H^{(l)} \) is given by:

\begin{equation}
    H^{(l)} = \text{FFN}(\text{SelfAttention}(H^{(l-1)})) + H^{(l-1)}
\end{equation}
where \( H^{(0)} = E \). The \(\text{SelfAttention}(\cdot)\) function computes contextual representations by attending to various parts of \( H^{(l-1)} \), while \(\text{FFN}(\cdot)\) is a position-wise feed-forward network.

The final hidden state vector \( \mathbf{H}_L \) of the last layer is mapped to a vocabulary space using a linear transformation, followed by a softmax function to obtain a probability distribution over the vocabulary: $\mathbf{z} = \mathbf{W_o} \mathbf{H}_L + \mathbf{b}$, where \( W_o \in \mathbb{R}^{d \times |V|} \) is the output layer weights and \( \mathbf{b} \) is the bias vector. This maps the \(d\)-dimensional outputs back to the vocabulary space \( |V| \), resulting in a probability distribution over potential output tokens. 

The probability of generating the next code token \( y \) is computed using the softmax function:
\begin{equation}
P(y) = \text{softmax}(\mathbf{z}) = \frac{\exp(\mathbf{z}_y)}{\sum_{j=1}^{|\mathcal{W}|} \exp(\mathbf{z}_j)}
\end{equation}
where \( \mathbf{z}_y \) is the score for the word \( y \) in the vocabulary, and \( |\mathcal{W}| \) is the size of the vocabulary.
\begin{equation}
P(Y|X) = \prod_{i=1}^{m}P(y_{i}|X, y_{1}, y_{2}, \cdots, y_ {i-1})
\end{equation}
The model generates a sequence of code tokens \( y_1, y_2, \ldots, y_m \) by sampling from the probability distributions \( P(y_1|X), P(y_2|X, y_1), \ldots, P(y_m|X, y_1, \ldots, y_{m-1}) \), where each token is conditioned on the previous tokens in the sequence.

\subsection{Model pruning}
Model pruning~\cite{liurethinking} is a well-established and efficient technique used for compressing models by reducing redundancy. Pruning methods are typically categorized as unstructured~\cite{sunsimple, mason2024makes} and structured~\cite{ma2023llm}.
In unstructured pruning, specific parameters, often individual weights or connections, are selectively removed without considering the model's internal structure. 
Unstructured pruning can be defined as
$\boldsymbol{W_{\text{pruned}}}=\boldsymbol{W}\odot M$, 
where $W$ represents the original weight matrix, $M$ is a binary mask indicating which weights to prune (0 for the pruned weights and 1 for the remaining weights), and $\odot$ denotes element-wise multiplication.

On the other hand, structured pruning offers a more systematic approach that preserves the overall architecture of the model. This method involves eliminating entire neurons, channels, or layers based on predefined criteria, such as the weight magnitudes or the importance of a neuron's contribution to the model's output.

%% file: sections/3method.tex
\section{Our Method}
\label{sec:method}

\begin{figure}[t]
    \centering
    \includegraphics[width=0.92\textwidth]{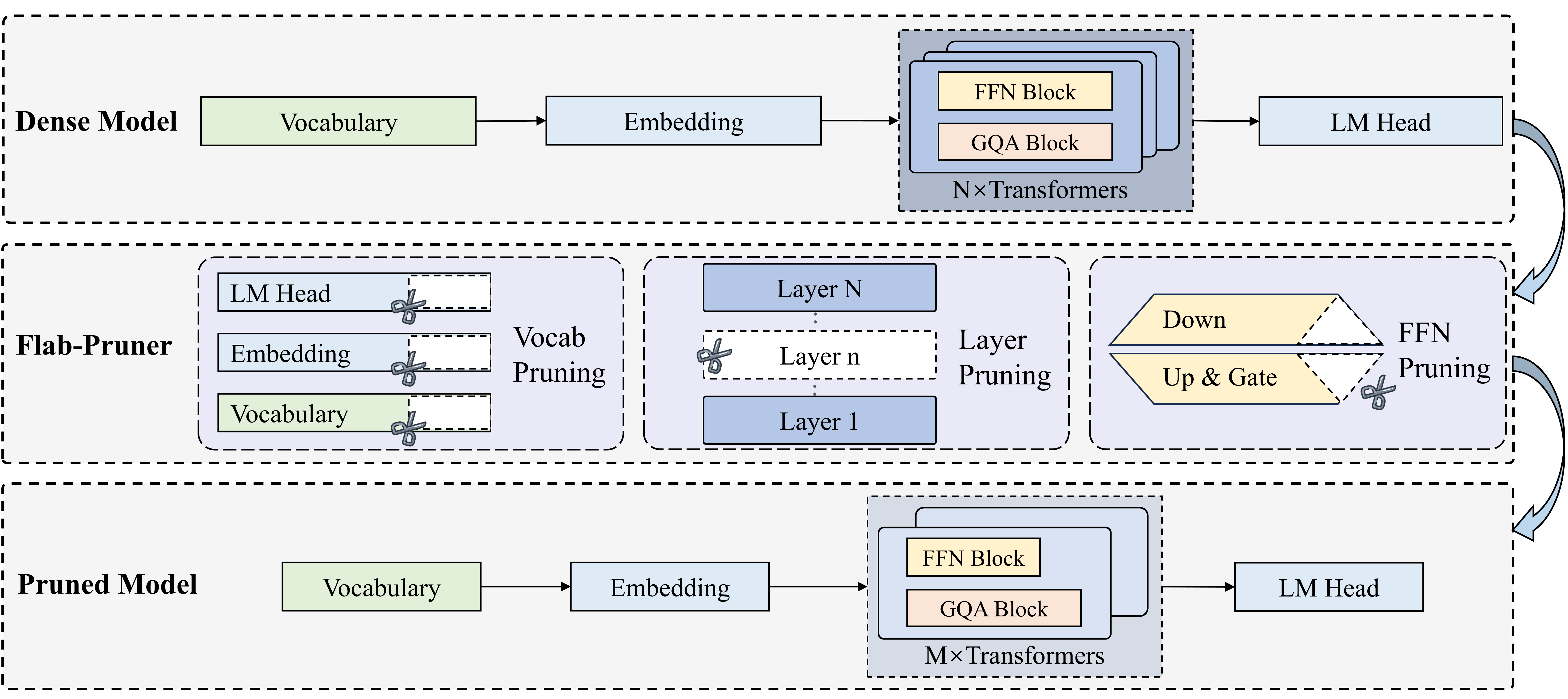}
    \caption{The framework of {\tool}}
    \vspace{-0.4cm}
    \label{fig:method}
\end{figure}

The workflow of {\tool} is shown in \figurename~\ref{fig:method}. There are three major pruning steps, i.e., vocab pruning, layer pruning and FFN pruning. 
For a given \yg{dense model}, we first perform vocab pruning, then layer pruning, and finally FFN pruning.

\subsection{Pruning Objective}
\label{subsec:pruning_objective}
To maintain the pruned model's performance, defining an appropriate pruning objective is crucial. 
In natural language processing, existing pruning methods typically focus on layer-level similarity. 
For instance, ShortGPT~\cite{men2024shortgpt} employs angle distance between layers, while Laco~\cite{yang2024laco} utilizes cosine similarity to identify and remove redundant layers.

However, for generative coding tasks, we argue that layer-level similarity alone is insufficient. 
The key challenge lies in ensuring that the pruned model generates code \(Y^{\prime}\) that remains consistent with the original model's output \(Y\).
Note that the model generates code through a probability distribution over the vocabulary, obtained by applying a linear transformation followed by a softmax function to the hidden representations \(H^{(L)}\) from the last layer.
Therefore, preserving the output probability distribution is critical for maintaining generation quality.

Based on this insight, we propose a pruning objective that minimizes the KL divergence between the output probability distributions of the pruned model (\(P_{\text{pruned}}\)) and the original model (\(P_{\text{original}}\)).
Formally, our objective can be expressed as:

\begin{equation}
    \label{eq:pruning_objective}
    \mathcal{L}_{\text{prune}} = \min_{\theta_p} D_{\text{KL}}(P(y|X;\theta_o) \| P(y|X;\theta_p))
    \end{equation}
    
    \noindent where \(\theta_o\) and \(\theta_p\) represent the parameters of the original and pruned models respectively, and \(D_{\text{KL}}\) denotes the Kullback-Leibler divergence between their output probability distributions:
    
    \begin{equation}
    \label{eq:kl_divergence}
    D_{\text{KL}}(P(y|X;\theta_o) \| P(y|X;\theta_p)) = \sum_{y \in \mathcal{V}} P(y|X;\theta_o) \log \frac{P(y|X;\theta_o)}{P(y|X;\theta_p)}
    \end{equation}

This objective ensures that the pruned model preserves the token generation probabilities of the original model, directly optimizing for the model's code generation behavior.
By focusing on the output probability distributions that determine code generation, our approach better aligns with the goal of maintaining task-specific performance in generative coding tasks.

\subsection{Vocabulary Pruning}

\begin{algorithm}[ht]
    \small
    \caption{Vocabulary Pruning Algorithm} 
    \label{algo:vocab_pruning_refactored}
    \KwIn{
        \\
        Original LLM $\mathcal{M}_{\text{ori}}$;\\
        Original Tokenizer $\mathcal{T}_{\text{ori}}$;\\
        Code DataSet $\mathcal{D}$;\\
    }
    \KwOut{
        \\
        Pruned LLM $\mathcal{M}_{\text{pru}}$;\\
        Pruned Tokenizer $\mathcal{T}_{\text{pru}}$;\\
    }
    \BlankLine
    $\mathcal{S} \gets \text{CollectTokens}(\mathcal{D}, \mathcal{T}_{\text{ori}})$\;
    $\mathcal{S} \gets \mathcal{S} \cup \mathcal{T}_{\text{ori}}[\text{``special~tokens"}]$\;
    \BlankLine
    $\mathcal{T}_{\text{pru}} \gets \text{PruneTokenizer}(\mathcal{T}_{\text{ori}}, \mathcal{S})$\;
    $\mathcal{M}_{\text{pru}} \gets \text{PruneModel}(\mathcal{M}_{\text{ori}}, \mathcal{T}_{\text{ori}}, \mathcal{T}_{\text{pru}}, \mathcal{S})$\;
    \BlankLine
    \Return $\mathcal{M}_{\text{pru}}, \mathcal{T}_{\text{pru}}$\;
    \BlankLine
    \SetKwFunction{FCollectTokens}{CollectTokens}
    \SetKwProg{Fn}{Function}{:}{}
    \Fn{\FCollectTokens{$\mathcal{D}, \mathcal{T}_{\text{ori}}$}}{
        $\mathcal{S} \gets \emptyset$\;
        \For{$code \in \mathcal{D}$}{
            \For{$token \in \mathcal{T}_{\text{ori}}.tokenize(code)$}{
                $\mathcal{S} \gets \mathcal{S} \cup \{token\}$\;
            }
        }
        \Return $\mathcal{S}$\;
    }
    \BlankLine
    \SetKwFunction{FPruneTokenizer}{PruneTokenizer}
    \Fn{\FPruneTokenizer{$\mathcal{T}_{\text{ori}}, \mathcal{S}$}}{
        $\textit{Vocab} \gets \emptyset$\;
        $\textit{Merges} \gets \emptyset$\;
        $id \gets 0$\;
        \For{$token \in \mathcal{S}$}{
            $\textit{Vocab}[token] \gets id$\;
            $id \gets id + 1$\;
        }
        \For{each $m \in \mathcal{T}_{\text{ori}}[\text{``merges"}]$}{
            \If{$\forall m[0], m[1], m[0] + m[1] \in \mathcal{S}$}{
                $\textit{Merges} \gets \textit{Merges} \cup \{m\}$\;
            }
        }
        $\mathcal{T}_{\text{pru}}[\text{``vocab"}] \gets \textit{Vocab}$\;
        $\mathcal{T}_{\text{pru}}[\text{``merges"}] \gets \textit{Merges}$\;
        \Return $\mathcal{T}_{\text{pru}}$\;
    }
    \BlankLine
    \SetKwFunction{FPruneModel}{PruneModel}
    \Fn{\FPruneModel{$\mathcal{M}_{\text{ori}}, \mathcal{T}_{\text{ori}}, \mathcal{T}_{\text{pru}}, \mathcal{S}$}}{
        $\textit{Embed} \gets \emptyset$\;
        $\textit{LM} \gets \emptyset$\;
        \For{$token \in \mathcal{S}$}{
            $\textit{Embed}[\mathcal{T}_{\text{pru}}[\text{``vocab"}][token]] \gets \mathcal{M}_{\text{ori}}[\text{``embed~tokens"}][\mathcal{T}_{\text{ori}}[\text{``vocab"}][token]]$\;
            $\textit{LM}[\mathcal{T}_{\text{pru}}[\text{``vocab"}][token]] \gets \mathcal{M}_{\text{ori}}[\text{``lm~head"}][\mathcal{T}_{\text{ori}}[\text{``vocab"}][token]]$\;
        }
        $\mathcal{M}_{\text{pru}}[\text{``embed~tokens"}] \gets \textit{Embed}$\;
        $\mathcal{M}_{\text{pru}}[\text{``lm~head"}] \gets \textit{LM}$\;
        \Return $\mathcal{M}_{\text{pru}}$\;
    }
\end{algorithm}

Vocabulary pruning, typically applied in tasks like text classification~\cite{chen2019large} and machine translation~\cite{nair2023blade}, involves removing these seldom-used tokens, thus reducing the overall model size without significantly affecting its capabilities. 

In the context of Code LLMs, many models have expanded their vocabularies to accommodate a wide range of programming languages. 
This expansion often leads to a large vocabulary size as the models strive to cover various syntaxes and language-specific terminologies. 
However, in practice, developers usually work with a limited set of programming languages, which means that a significant portion of the vocabulary is rarely used. 
This observation leads to the hypothesis that the absence of these infrequent tokens might not substantially impact the model's ability to capture the nuances of any particular language effectively~\cite{yang2022textpruner}.

\noindent\textbf{Formalization.} 
Formally, given a vocabulary \( V \) and a usage statistic \( U \), the pruned vocabulary \( V' \) is defined as:
\begin{equation}
    V' = \{v \in V \mid U(v) > \tau\}
\end{equation}
where \(\tau\) is a threshold value representing the minimum frequency a token must have to remain in the vocabulary. 
Consequently, the embedding matrix \( W_e \in \mathbb{R}^{|V| \times d} \) is reduced to \( W'_e \in \mathbb{R}^{|V'| \times d} \), removing unnecessary token embeddings and reducing model complexity.
Meanwhile, the output layer weights \( W_o \in \mathbb{R}^{d \times |V|} \) is also reduced to \( W'_o \in \mathbb{R}^{d \times } |V'|\).

\noindent\textbf{Algorithms.} 
In the realm of LLMs, the byte-level BPE algorithm~\cite{wang2020neural} is widely adopted for tokenization, as seen in models such as LLAMA~\cite{touvron2023llama} and QWen~\cite{bai2023qwen}.
The BPE consists of two key elements: the vocabulary, which is a set of tokens that the model identifies to interpret data, and merge rules, which guide the combination of token pairs into complex structures. 

Algorithm~\ref{algo:vocab_pruning_refactored} provides a detailed pseudo-code description of the vocabulary pruning process, structured into modular steps for enhanced clarity and maintainability. 
The process begins with the collection of all unique tokens from a given code dataset using the original tokenizer, as implemented in the `CollectTokens' function (Lines 6-11). 
This set of tokens, $\mathcal{S}$, also includes essential special tokens (Line 3). The algorithm then moves to prune the tokenizer via the `PruneTokenizer' function, where a streamlined vocabulary and updated merge rules are created based only on the tokens present in the collected set (Lines 12-24). 
Subsequently, the `PruneModel' function handles the pruning of the language model itself, retaining only the embeddings and language model output layers corresponding to the preserved tokens (Lines 25-33). 
The final output, consisting of a reduced and optimized version of the original model and tokenizer, ensures the preservation of core functional capabilities while eliminating redundant elements (Lines 4-5).

\noindent\textbf{Analysis.} The computational expense associated with vocabulary pruning is primarily attributed to the tokenization process, which can vary from a few seconds to several minutes contingent upon the size of the corpus.


\subsection{Layer Pruning}
Layer pruning involves the removal of entire layers to reduce its depth and computation, which is theoretically justified by the observed redundancy~\cite{dalvi2020analyzing} within deep neural networks and the notable similarity~\cite{gromov2024unreasonable} across layers.
In NLP, studies~\cite{men2024shortgpt, chen2024compressing, song2024sleb, kim2024shortened} reveal that the performance of LLMs remains largely intact even after the removal of many layers, indicating an underlying redundancy that pruning can exploit.

\begin{algorithm}[t]
    \small
    \caption{Layer Pruning Algorithm}
    \label{algo:layer_pruning_refactored}
    \KwIn{
        \\
        Original LLM $\mathcal{M}_{\text{ori}}$;\\
        Code Generation Datasets $\mathcal{D}$;\\
        Number of layers to prune $k$;\\
    }
    \KwOut{
        \\
        Pruned LLM $\mathcal{M}_{\text{pru}}$;\\
    }
    \BlankLine
    $\mathcal{D}_{\text{correct}} \gets \text{FilterCorrectSamples}(\mathcal{D}, \mathcal{M}_{\text{ori}})$\;
    $P_{\text{original}} \gets \mathcal{M}_{\text{ori}}(\mathcal{D}_{\text{correct}})$\;
    $\mathcal{M}_{\text{pru}} \gets \mathcal{M}_{\text{ori}}$\;
    \BlankLine
    \While{number of layers removed $<$ $k$}{
        $best\_layer, best\_score \gets \text{FindBestLayerToPrune}(\mathcal{M}_{\text{pru}}, H^{(L)}_{\text{original}})$\;
        $\mathcal{M}_{\text{pru}} \gets \text{RemoveLayer}(\mathcal{M}_{\text{pru}}, best\_layer)$\;
    }
    \BlankLine
    \Return $\mathcal{M}_{\text{pru}}$\;
    \BlankLine
    \SetKwFunction{FFilterCorrectSamples}{FilterCorrectSamples}
    \SetKwProg{Fn}{Function}{:}{}
    \Fn{\FFilterCorrectSamples{$\mathcal{D}, \mathcal{M}_{\text{ori}}$}}{
        $\mathcal{D}_{\text{correct}} \gets \emptyset$\;
        \For{$sample \in \mathcal{D}$}{
            \If{$\text{isCorrectlyGenerated}(sample, \mathcal{M}_{\text{ori}})$}{
                $\mathcal{D}_{\text{correct}} \gets \mathcal{D}_{\text{correct}} \cup \{sample\}$\;
            }
        }
        \Return $\mathcal{D}_{\text{correct}}$\;
    }
    \BlankLine
    \SetKwFunction{FFindBestLayerToPrune}{FindBestLayerToPrune}
    \Fn{\FFindBestLayerToPrune{$\mathcal{M}_{\text{pru}}, \yg{P_{\text{original}}}$}}{
        $best\_score \gets \yg{\infty}$\;
        $best\_layer \gets \text{None}$\;
        \For{each layer $l$ in $\mathcal{M}_{\text{pru}}$}{
            $\mathcal{M}_{\text{temp}} \gets \text{RemoveLayer}(\mathcal{M}_{\text{pru}}, l)$\;
            $P_{\text{pruned}} \gets \mathcal{M}_{\text{temp}}(\mathcal{D}_{\text{correct}})$\;
            $score \gets \text{kl\_divergence}(P_{\text{pruned}}, P_{\text{original}})$\;
            \If{score $<$ best\_score}{
                $best\_score \gets score$\;
                $best\_layer \gets l$\;
            }
        }
        \Return $best\_layer, best\_score$\;
    }
\end{algorithm}

\noindent\textbf{Algorithms.}
Our layer pruning approach diverges from existing methods that typically rely on layer similarity or language modeling capabilities as the primary criterion for pruning. 
We assert that the primary purpose of pruning should be to ensure that the model continues to deliver satisfactory performance, which we discussed in Section~\ref{subsec:pruning_objective}. 

Furthermore, existing methods~\cite{ma2023llm,men2024shortgpt,gromov2024unreasonable,razzhigaev2024your,song2024sleb} primarily focus on one-shot removal strategies, which determine the redundancy of each layer in a single computation pass and select the top$\_k$ layers or a contiguous block of $k$ layers for removal. 
These methods are straightforward but may not account for the dynamic changes in the model's state following the removal of layers.
Beyond one-shot removal used in existing studies, our approach deploys an iterative process to meticulously evaluate each layer's contribution to the semantic representation of the final model output.

Algorithm~\ref{algo:layer_pruning_refactored} provides the pseudo-code to describe the detailed layer pruning process. 
The algorithm is designed to iteratively prune the layers of a LLM to reduce its complexity while maintaining performance. 
It starts by filtering the code generation dataset to identify samples for which the model correctly generates code, as detailed in the `FilterCorrectSamples' function (Lines 8-13). 
The main pruning loop (Lines 4-6) continues until the specified number of layers, $k$, have been pruned. 
In each iteration, the `FindBestLayerToPrune' function is used to determine the most dispensable layer by calculating the KL divergence between the probability distribution of the original and pruned models(Lines 14-24). 
Once identified, this layer is permanently removed from the model using `RemoveLayer'. 
The process returns a pruned version of the original model, $\mathcal{M}_{\text{pru}}$.

\noindent\textbf{Analysis.}
The computational cost associated with our layer pruning algorithm primarily stems from the iterative evaluation of layer removal and the subsequent similarity calculations between the pruned and original models' final layer representations.


\subsection{FFN Pruning}

FFN pruning is based on the observation that not all neurons in transformer models are equally vital~\cite{zhu2023survey}. 
Each layer in Transformer is composed of a GQA module and an FFN module, the GQA's inherent characteristics necessitate a specific ratio between $key\_value\_heads$ and $attention\_heads$. Pruning any can lead to a significant degradation in model performance.
Given this constraint, our focus shifts to the intermediate size of the FFN, which represents the expansion of the model's representational capacity and presents a more flexible target for pruning. 

\noindent\textbf{Algorithm.}
To streamline the FFN pruning process and minimize its computational overhead, we have devised a set of four heuristic rules to determine which neurons within the FFN to eliminate:
\begin{itemize}
    \item \textbf{Top-K Neurons.} Retain the first K neurons, which are hypothesized to be the most influential.
    \item \textbf{Bottom-K Neurons.} Retain the last K neurons, assuming they may carry critical information not present in the initial neurons.
    \item \textbf{Middle-K Neurons.} Retain the middle K neurons, which might be less prone to noise and more stable in their contribution.
    \item \textbf{Random Sampling.} Randomly select K neurons to be retained, introducing an element of stochasticity to the pruning process.
\end{itemize}

Following these heuristics, we generate a structured pruning mask that guides the pruning process.
Finally, we still choose the best rule through computing the similarity.

\noindent\textbf{Analysis.}
Our approach to FFN pruning leverages heuristic rules, which significantly reduces the computational cost. The time required for this process is merely a matter of minutes, making it a highly efficient method for model optimization.

\subsection{Performance Recovery}
After pruning some model parameters, it is crucial to implement effective strategies to recover the model's performance. 
For performance recovery, we fine-tune the pruned model on the training split of our CodeHarmony dataset (detailed in Section 4.3), which contains 15,800 samples for code generation and CoT tasks and 100,706 samples for output prediction tasks.
The following pseudocode describes our proposed Performance Recovery Algorithm.

\begin{algorithm}[t]
    \small
    \caption{Performance Recovery Algorithm}
    \label{algo:performance_recovery}
    \KwIn{
        \\
        Original LLM $\mathcal{M}_{\text{ori}}$;\\
        Pruned LLM $\mathcal{M}_{\text{pru}}$;\\
        Training Dataset with Test Case Set $\mathcal{D}=[x, y ,T]$;\\
    }
    \KwOut{
        \\
        Recovered LLM $\mathcal{M}_{\text{rec}}$;\\
    }
    \BlankLine

    \BlankLine
    \For{sample $s \in \mathcal{D}$}{
        $original\_code \gets \mathcal{M}_{\text{ori}}.\text{generate}(s.x)$\;
        \If{TestPass(original\_code, $s.T$)}{
            $s.y \gets original\_code$\;
            $\mathcal{D}$.updates$(s)$\;
        }
    }
    
    \BlankLine
    $\mathcal{M}_{\text{rec}} \gets \text{Train}(\mathcal{M}_{\text{pru}}, \mathcal{D})$\;
    
    \Return $\mathcal{M}_{\text{rec}}$\;
    \BlankLine
    
    \SetKwFunction{FTestPass}{TestPass}
    \SetKwProg{Fn}{Function}{:}{}
    \Fn{\FTestPass{$code, T$}}{
        \For{test case $t \in T$}{
            \If{not isPass(code, $t$)}{
                \Return \text{False}\;
            }
        }
        \Return \text{True}\;
    }
\end{algorithm}

Algorithm~\ref{algo:performance_recovery} presents the pseudocode for the Performance Recovery process. 
The algorithm aims to improve the performance of the pruned model by leveraging the code generation capabilities of the original model.
The process begins by iterating through the provided training dataset $\mathcal{D}$, which consists of samples including input data, expected outputs, and associated test cases (Lines 2-5). 
For each sample, the algorithm uses the original model $\mathcal{M}_{\text{ori}}$ to generate code based on the input data. 
If the generated code passes the corresponding test cases, the expected output in the dataset is replaced with this generated code.

By replacing outputs with semantically correct code generated by the original model, the overall quality of the dataset is consistently high, ensuring the training data is always reliable.
Furthermore, the pruned model is able to fit the training data more efficiently, reducing loss more quickly and effectively during training. 
This enhances the convergence speed and helps the pruned model achieve better performance faster.

Finally, the pruned model $\mathcal{M}_{\text{pru}}$ is retrained on the updated dataset. 
This training aims to equip the pruned model with the necessary knowledge to handle previously challenging scenarios. 
The outcome is a recovered version of the pruned model, denoted as $\mathcal{M}_{\text{rec}}$.
We employ the LoRA technique~\cite{hulora} for a refined and efficient post training of the pruned model, which is the most efficient alternative to full-parameter fine-tuning~\cite{zhuo2024astraios, weyssow2023exploring, liu2023empirical}.
It is worth noting that LoRA does not increase the number of parameters of the model.
The time cost for post training is about a few hours, which is contingent upon the size of the dataset.

%% file: sections/4setup.tex
\section{Experiment Setup}
\label{sec:setup}

To assess the effectiveness of {\tool}, we design the following three research questions (RQs):

\begin{description}
\item[\textbf{RQ1}] (Performance Comparison) 
This RQ is designed to evaluate the performance of {\tool} compared to dense models and other structured pruning methods.
Furthermore, we investigate the impact of three components proposed in {\tool} and conducts a hyperparameter analysis.

\item[\textbf{RQ2}] (Efficiency Comparison)
This RQ is designed to examine the resource deployment efficiency of {\tool} compared to dense models.

\item[\textbf{RQ3}] (Robustness Analysis)
This RQ is designed to assess the robustness of pruned models compared to dense models, particularly under various prompt perturbations. 

\end{description}

\subsection{Subject Models}
In our research, we select three state-of-the-art CodeLLMs, i.e., CodeQwen-1.5 and its two variants (NxCode and CodeSlerp), which are top-performing 7B-parameter models on the HumanEval leaderboard. 
These models are chosen for their strong performance and all three models share the same transformer-based architecture, making them ideal candidates for our comparative study of pruning techniques.

\subsection{Downstream Tasks}
In our research, we focus on the generation tasks in code intelligence. We delve into the following three key tasks, each with its unique set of challenges and methodologies:
\begin{itemize}
    \item \textbf{Code Generation.} This task~\cite{chen2021evaluating} involves the generation of code snippets from a given natural language description and signature, utilizing a zero-shot learning strategy. 
    
    \item \textbf{CoT Generation.} This task~\cite{yang2023chain} involves the generation of a chain of thought leading to the solution, given a natural language description and signature. It is executed through a one-shot learning approach, where the model learns from a single example.

    \item \textbf{Output Prediction.} This task~\cite{gucruxeval} focuses on predicting the output of a code snippet without execution, employing a two-shot learning approach. The model must deduce the expected output based on the code's logic, given the code and its corresponding input.
\end{itemize}

\subsection{Datasets}
For the code generation task, we focus on function-level code generation, primarily using the widely-used HumanEval\footnote{\url{https://huggingface.co/datasets/openai/openai_humaneval}} and OpenEval\footnote{\url{https://huggingface.co/datasets/NTUYG/openeval}} for broader evaluation.
For the CoT generation task, we employ the HumanEval-CoT and OpenEval-CoT datasets, developed by Yang et al.~\cite{yang2023chain}.
For the output prediction task, we select the CRUX dataset\footnote{\url{https://huggingface.co/datasets/cruxeval-org/cruxeval}}, introduced by Meta~\cite{gu2024cruxeval}.

\begin{table}[tbp]
    \centering
    \caption{Statistical information of our selected tasks and corresponding datasets, existing datasets are limited in number of samples and have no training data.}
    \begin{tabular}{ccccc}
    \toprule
    Task & Dataset & Train & Valid & Test\\ 
    \midrule
    Code Genaration & HumanEval & -  & -  & 164\\
    (Zero-Shot) & OpenEval & -  & -  & 178\\
    & CodeHarmony & 15,800 & 200 & 153\\
    \midrule
    CoT Genaration & HumanEval & -  & -  & 164\\
    (One-Shot) & OpenEval & -  & -  & 178\\
    & CodeHarmony & 15,800 & 200 & 153\\
    \midrule
    Output Prediciton & Crux-O & -  & -  & 800\\
    (Two-Shot) & CodeHarmony & 100,706 & -  & 452\\
    \bottomrule
    \end{tabular}
    \label{tab:dataset} 
\end{table}


Moreover, to address the limitations of existing datasets in terms of scale, we introduce CodeHarmony, a comprehensive benchmark dataset for code intelligence evaluation. The construction of CodeHarmony follows a systematic three-step process:

\begin{enumerate}
    \item \textbf{Data Collection:} We aggregate function-level Python code from well-established open-source repositories, primarily the Evol dataset~\cite{luowizardcoder} and the OSS dataset~\cite{wei2024magicoder}. Using carefully designed regular expressions, we extract syntactically valid and self-contained functions to ensure code quality.
    
    \item \textbf{Test Case Generation:} To validate the semantic correctness of the collected code, we implement a hybrid verification approach. We leverage state-of-the-art language models (GPT-4o and Gemini) to automatically generate diverse test cases.
Specifically, we follow Beau et al.~\cite{beau2024codeinsight} to construct three test cases for each function to ensure the code's correctness.
    
    \item \textbf{Chain-of-Thought Integration:} Inspired by recent advances in multi-agent alignment techniques~\cite{yang2023chain}, we augment the dataset with CoT, which provides intermediate reasoning steps, enhancing the dataset's utility for evaluating models' reasoning capabilities in code understanding and generation tasks.
\end{enumerate}

The resulting dataset comprises a diverse collection of code samples with corresponding test cases and CoT. Table~\ref{tab:dataset} presents detailed statistics of CodeHarmony, demonstrating its comprehensive coverage and suitability for evaluating various aspects of code intelligence models.

\subsection{Evaluation Metrics}
To assess the efficacy of the models in these tasks, we have adopted the following evaluation metrics.  

\begin{itemize}
	
\item\textbf{Pass@1} quantifies the percentage of generated code snippets that successfully pass the associated test cases~\cite{chen2021evaluating}.

\item \textbf{BLEU} evaluates lexical similarity overlap between texts using n-gram comparison~\cite{papineni2002bleu}. We use BLEU-4 to assess the quality of the generated CoTs.

\item  \textbf{Exact Match (EM)} metric evaluates the precision by determining if the generated output exactly matches the expected output.
\end{itemize}

\subsection{Implementation details.}
For FFN pruning, we remove the 256 neurons of each layer.
For layer pruning, we remove the 4 layers of the model.
For vocabulary pruning, we reduce the vocabulary size from 92,416 to 17,176.
Based on the above settings, we reduce the model's parameter size from 7.3 billion to 5.7 billion, a reduction of 22\%.

For the model inference, we use greedy decoding method to calculate the Pass@1, BLEU and EM scores.
For the remaining hyperparameters, we have meticulously tuned the following settings: a learning rate of 5e-4, a LoRA rank of 64, and a LoRA alpha of 32.
Our experiments, using PyTorch and Transformers, were run on a system with an Intel Xeon Silver 4210 CPU and a GeForce RTX 3090 GPU. The model pruning and post-training took approximately 6 hours to complete.


%% file: sections/5results.tex
\section{Experiment Result}
\label{sec:result}

\subsection{RQ1: Performance Comparison} \label{subsect:rq1}

\begin{table*}[htbp]
\centering
\caption{Performance comparison of {\tool} and other baselines, where HE represents HumanEval, OE represents OpenEval, and CH represents CodeHarmony. The best results (compared with other pruning methods) are highlighted in bold. The best results (compared with the dense model) are highlighted in gray.
FT means traditional fine-tuning, and PT means post-training we proposed.}
\resizebox{\textwidth}{!}{
\begin{tabular}{ccccccccccccc}
\toprule
\multirow{2}{*}{Model} & \multirow{2}{*}{Method} & \multicolumn{4}{c}{Code Generation(Pass@1)} & \multicolumn{4}{c}{CoT Generation(BLEU-4)} & \multicolumn{3}{c}{Output Prediciton(EM)} \\ 
\cmidrule(lr){3-6} \cmidrule(lr){7-10} \cmidrule(lr){11-13} 
& & HE & OE & CH & Avg. & HE-CoT & OE-CoT & CH-CoT & Avg. & Crux-O & CH-O & Avg. \\
\midrule
\multirow{9}{*}{CodeQwen} & Dense & 77.44 & \cellcolor{\best}42.13 & 60.78 & 60.12 & 33.95 & \cellcolor{\best}41.14 & 23.81 & \cellcolor{\best}32.97 & 37.13 & 77.43 & 57.28\\
\cmidrule(lr){2-13}
& ShortGPT & 42.68 & 20.79 & 41.83 & 35.10 & 15.92 & 19.02 & 11.41 & 15.45 & 27.63 & 59.29 & 43.46\\
& UIDL & 0 & 0 & 0 & 0 & 0 & 0 & 0 & 0  & 0 & 0 & 0 \\
& Linearity & 0 & 0 & 0.65 & 0.22 & 0 & 0 & 0 & 0 & 0 & 0 & 0\\
& SLEB & 20.73 & 17.42 & 49.67 & 29.27 & 1.72 & 5.58 & 2.92 & 3.41 & 18.75 & 42.70 & 30.73\\
& LLM-pruner & 15.85 & 10.67 & 37.25 & 21.25 & 3.88 & 8.33 & 3.99 & 5.40 & 16.50 & 25.44 & 20.97\\
& {\tool} & \textbf{75.00} & \textbf{37.64} & \textbf{64.05} & \textbf{58.90} & \textbf{31.42} & \textbf{35.88} & \textbf{20.80} & \textbf{29.37} & \textbf{31.75} & \textbf{70.58} & \textbf{51.17}\\
\cmidrule(lr){2-13} 
& {\tool} w FT & 76.22 & 36.52 & 66.63 & 59.79 & 32.17 & 34.65 & 25.80 & 30.87 & 40.38 & 77.10 & 58.74\\
& {\tool} w PT & \cellcolor{\best}78.05 & 39.89 & \cellcolor{\best}67.97 & \cellcolor{\best}61.97 & \cellcolor{\best}34.37 & 34.67 & \cellcolor{\best}26.60 & 31.88 & \cellcolor{\best}40.63 & \cellcolor{\best}77.43 & \cellcolor{\best}59.03\\
\midrule
\multirow{9}{*}{NxCode} & Dense & 77.44 & \cellcolor{\best}41.57 & 62.09 & 60.37 & 32.87 & \cellcolor{\best}39.90 & 24.15 & \cellcolor{\best}32.31 & 37.00 & \cellcolor{\best}77.43 & 57.22\\
\cmidrule(lr){2-13}
& ShortGPT & 40.85 & 22.47 & 38.56 & 33.96 & 19.03 & 20.00 & 12.53 & 17.19 & 28.75 & 59.51 & 44.13\\
& UIDL & 0 & 0 & 0 & 0 & 0 & 0 & 0 & 0  & 0 & 0 & 0 \\
& Linearity & 0 & 0 & 0.65 & 0.22 & 0 & 0 & 0 & 0 & 0 & 0 & 0\\
& SLEB & 20.12 & 17.98 & 47.71 & 28.60 & 2.42 & 5.88 & 3.37 & 3.89 & 18.38 & 41.15 & 29.77\\
& LLM-pruner & 17.07 & 9.55 & 36.60 & 21.07 & 3.31 & 8.96 & 4.13 & 5.47 & 15.00 & 23.23 & 19.12\\
& {\tool} & \textbf{74.39} & \textbf{35.96} & \textbf{64.05} & \textbf{58.13}  & \textbf{31.03} & \textbf{35.31} & \textbf{21.76} & \textbf{29.37} & \textbf{32.63} & \textbf{70.58} & \textbf{51.61}\\
\cmidrule(lr){2-13} 
& {\tool} w FT & 75.51 & 38.20 & 62.36 & 58.69 & 30.86 & 34.88 & 24.14 & 29.96 & 40.13 & 77.10 & 58.62\\
& {\tool} w PT & \cellcolor{\best}81.71 & 39.33 & \cellcolor{\best}64.05 & \cellcolor{\best}61.70 & \cellcolor{\best}32.96 & 37.40 & \cellcolor{\best}24.71 & 31.69& \cellcolor{\best}40.63 & 77.21 & \cellcolor{\best}58.92\\
\midrule
\multirow{9}{*}{CodeSlerp} & Dense & 77.44 & \cellcolor{\best}43.26 & 61.44 & 60.71 & 32.18 & \cellcolor{\best}40.49 & 24.51 & \cellcolor{\best}32.39 & 37.25 & \cellcolor{\best}76.77 & \cellcolor{\best}57.01\\
\cmidrule(lr){2-13}
& ShortGPT & 41.46 & 21.35 & 39.61 & 34.14 & 18.03 & 20.00 & 12.53 & 16.85 & 29.38 & 60.00 & 44.69\\
& UIDL & 0 & 0 & 0 & 0 & 0 & 0 & 0 & 0  & 0 & 0 & 0 \\
& Linearity & 0 & 0 & 0.65 & 0.22 & 0 & 0 & 0 & 0 & 0 & 0 & 0\\
& SLEB & 21.34 & 18.54 & 48.37 & 29.42 & 2.12 & 5.22 & 3.12 & 3.49 &18.44 & 41.25 & 29.85\\
& LLM-pruner & 16.89 & 11.35 & 38.60 & 22.28 & 3.65 & 9.25 & 5.11 & 6.00 & 15.25 & 24.25 & 19.75\\
& {\tool} & \textbf{75.00} & \textbf{35.96} & \textbf{64.05} & \textbf{58.34} & \textbf{31.48} & \textbf{35.96} & \textbf{21.83} & \textbf{29.76} & \textbf{32.25} & \textbf{70.35} & \textbf{51.30}\\
\cmidrule(lr){2-13} 
& {\tool} w FT & 73.17 & 37.20 & 65.36 & 58.58 & 31.66 & 34.03 & 25.81 & 30.50 & 37.50 & 75.60 & 56.55\\
& {\tool} w PT & \cellcolor{\best}78.66 & 37.64 & \cellcolor{\best}66.01 & \cellcolor{\best}60.77 & \cellcolor{\best}32.27 & 37.17 & \cellcolor{\best}25.96 & 31.80 & \cellcolor{\best}37.63 & 75.66 & 56.65\\
\bottomrule
\end{tabular}
}
\label{tab:rq-1} 
\end{table*}

One of the primary objectives of this study is to evaluate the effectiveness of {\tool} in reducing model size while preserving the performance.
To this end, we design a series of experiments to compare the performance of {\tool} against other structured pruning methods as well as the original dense models. 
We consider five representative pruning methods, including ShortGPT, UIDL, Linearity, SLEB and LLM-pruner as the baselines. Each of these methods has a unique focus and explores different aspects of model pruning.
\yg{For each baseline, we control the pruning ratio to be 20\% (slightly lower than our method's 22\%) to ensure a fair comparison across all methods.}

\begin{itemize}
    \item \textbf{ShortGPT}~\cite{men2024shortgpt}  prunes layers guided by the \textit{cosine similarity} between layer representations.
    
    \item \textbf{UIDL}~\cite{gromov2024unreasonable} prunes layers by evaluating the \textit{angular distance} between layer representations.
    
    \item \textbf{Linearity}~\cite{razzhigaev2024your} prunes layers by evaluating the \textit{linear relationship} of layer representations.
    
    \item \textbf{SLEB}~\cite{song2024sleb} prunes layers that have the least impact on the model by calculating the \textit{perplexity metric}.
    
    \item \textbf{LLM-pruner}~\cite{ma2023llm} prunes layers that have the least impact on the model by utilizing gradients derived from \textit{Taylor's formula}.
\end{itemize}

\underline{\textbf{(1) Compared with single component pruning methods:}}
Our empirical results, shown in Table~\ref{tab:rq-1}, demonstrate the performance of {\tool} across various tasks. 
Without post-training, {\tool} achieves the highest performance in the code generation task.
Taking HumanEval as an example, the Pass@1 of {\tool} on the CodeQwen model is 75.00\%, while the best Pass@1 of other baseline methods is 42.68\%, compared to the relative performance improvement of 75.73\%.
In the CoT generation task, {\tool} also performs better, 
with a BLEU-4 of 31.42\% on the CodeQwen model, while the best BLEU-4 of other baseline methods is 15.92\%, compared to the relative performance improvement of 97.98\%.
For the code output prediction task, {\tool} also outperforms other methods.
Taking Crux-O as an example, the EM of {\tool} on the CodeQwen model is 31.75\%, while the best EM of other baseline methods is 27.63\%, compared to the relative performance improvement of 14.93\%.

\begin{figure}[htbp]
    \centering
    \subfigure[Results on CodeQwen model]{\includegraphics[width=0.3\textwidth]{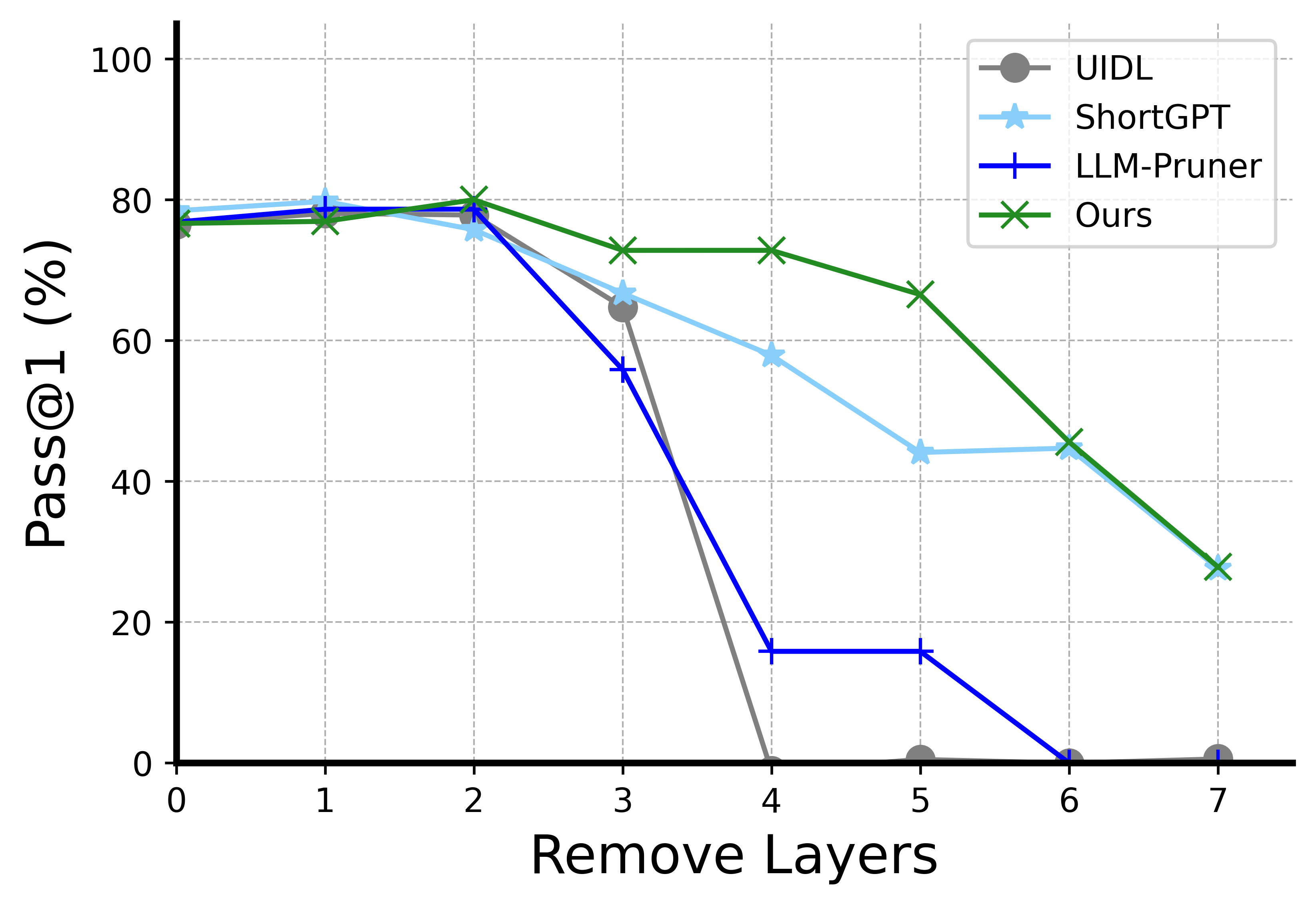}
    \label{fig:CodeQwen}
    }
    \subfigure[Results on Nxcode model]{\includegraphics[width=0.3\textwidth]{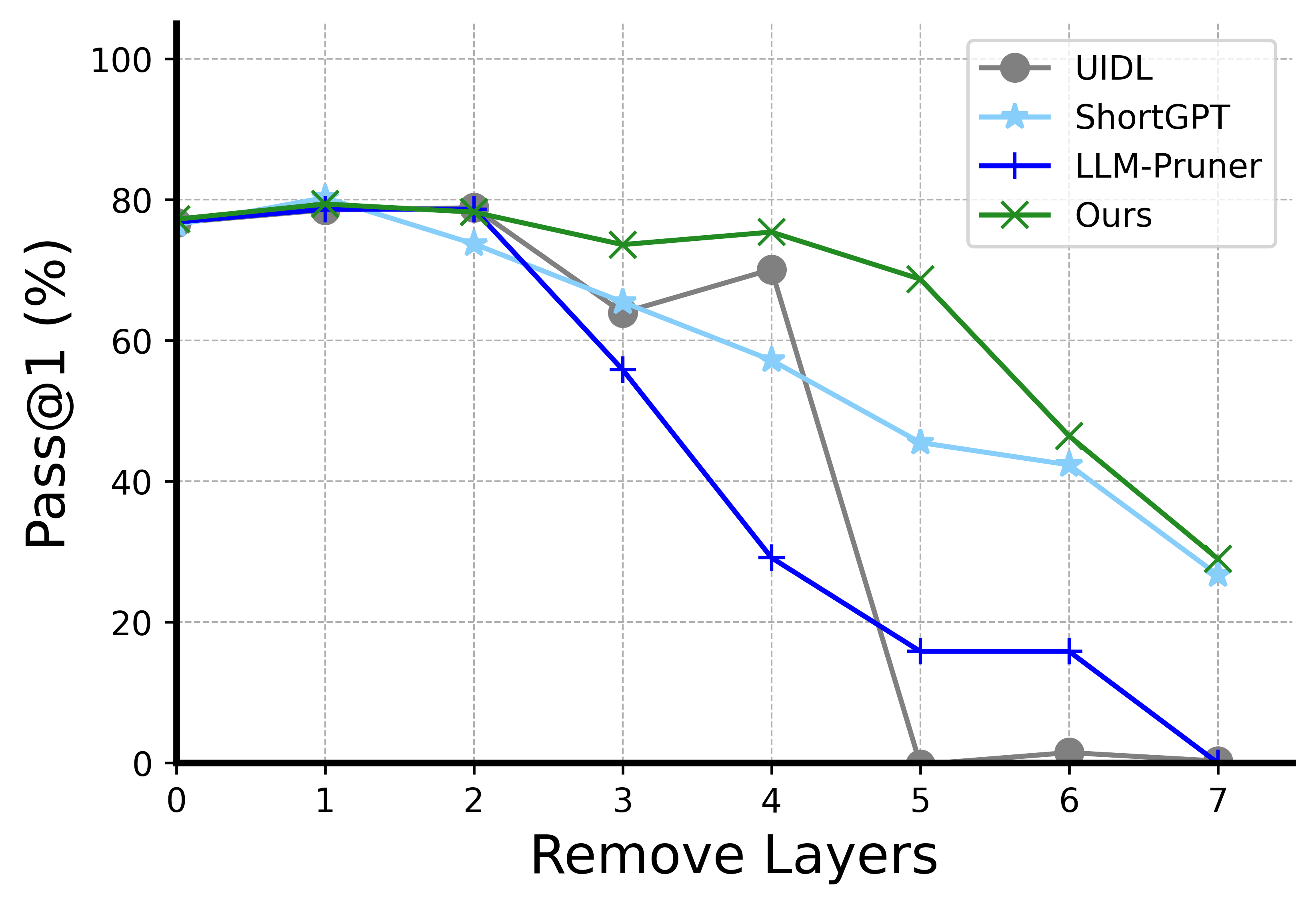}
    \label{fig:Nxcode}
    }
    \subfigure[Results on CodeSlerp model]{\includegraphics[width=0.3\textwidth]{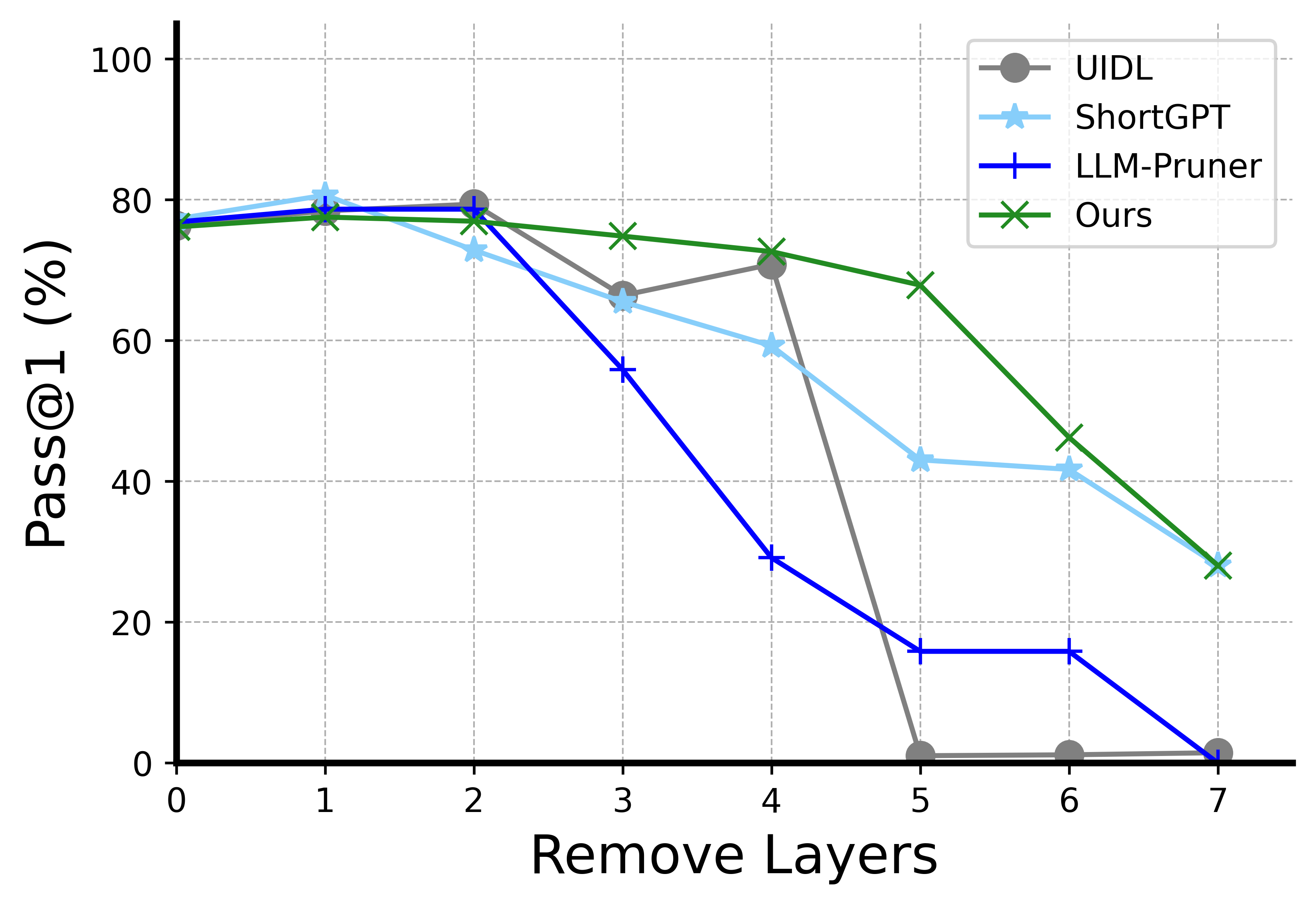}
    \label{fig:CodeSlerp}
    }
    \caption{\yg{Performance comparison of different layer pruning models with different number of layers pruned.}}
    \label{fig:layer}
    \vspace{-0.3cm}
\end{figure}

\underline{\textbf{(2) Compared with other pruning objectives:}}
As shown in Figure~\ref{fig:layer}, we compare the effect of pruning different number of layers on the model performance and find that the more layers are pruned, the more the model performance decreases. 
Furthermore, we conduct an in-depth analysis of our proposed pruning objectives in Layer Pruning.
\yg{We conducted comprehensive comparisons between our KL divergence-based approach and several alternative pruning objectives. 
When examining the impact of the progressively increasing  number of pruned layers, we observe that with minimal pruning (1-2 layers), all methods maintain performance comparable to the original model. 
However, as more layers are removed, significant differences emerge between pruning objectives. 
The angular distance-based method (UIDL) deteriorates most dramatically, with pass@1 metrics dropping from above 60\% to 0\% when pruning beyond a certain threshold. 
Both Taylor formula-based gradients (LLM-Pruner) and cosine similarity-based methods (ShortGPT) show substantial performance degradation as pruning increases, while our KL divergence approach maintains higher performance with a more gradual decline. 
When pruning ratios become extremely high, all methods eventually experience severe performance drops, though our approach consistently degrades more gracefully.
These results demonstrate that KL divergence  captures layer importance for code intelligence tasks more effectively than both gradient-based and representation similarity-based methods.}

These performance differences are primarily attributed to the inherent limitations in the layer selection and importance assessment strategies of other methods. 
SLEB’s reliance on perplexity metrics may not fully correspond to code functionality accuracy. 
ShortGPT, UIDL, and Linearity’s isolated layer evaluations fail to capture the collective impact of layers on overall model performance. 
Additionally, LLM-Pruner’s dependence on back propagation for gradient computation introduces unnecessary computational overhead for code intelligence tasks.
This suggests that our proposed layer pruning strategy can better capture the importance of the each layer, thus alleviating the problem of inconsistent pruning objectives mentioned in Section~\ref{sec:intro} and better guiding model pruning.

\underline{\textbf{(3) Strengths in our proposed performance recovery strategy:}}
Compared to traditional fine-tuning method, our proposed strategy achieves better performance recovery.
As shown in Table~\ref{tab:rq-1}, after post-training, {\tool} achieves the best performance across all tasks in three models.
Taking HumanEval as an example, the Pass@1 of {\tool} on the CodeQwen model is 78.05\%, while the best Pass@1 of traditional fine-tuning method is 76.22\%, compared to the relative performance improvement of 2.40\%.

\begin{table*}[tbp]
    \centering
    \caption{Ablation Comparison of {\tool} with different pruning components.}
    \resizebox{\textwidth}{!}{
    \begin{tabular}{cccccccccc}
    \toprule
    \multirow{2}{*}{Model} & \multirow{2}{*}{Method} & \multicolumn{3}{c}{Code Generation(Pass@1)} & \multicolumn{3}{c}{CoT Generation(BLEU-4)} & \multicolumn{2}{c}{Output Prediciton(EM)} \\ 
    \cmidrule(lr){3-5} \cmidrule(lr){6-8} \cmidrule(lr){9-10} 
    & & HE & OE & CH & HE-CoT  & OE-CoT & CH-CoT & Crux-O & CH-O\\
    \midrule
    \multirow{10}{*}{CodeQwen} & Dense & 77.44 & 42.13 & 60.78 & 33.95 & 41.14 & 23.81 & 37.13 & 77.43\\
    & + Vocab Pruning & 77.44 & 42.70 & 62.09 & 33.85 & 41.44 & 22.21 & 36.50 & 75.66\\
    \cmidrule(lr){2-10}
    & + FFN (Top-K) & 74.39 & 37.64 & 64.05 & 31.42 & 35.88 & 20.80 & 31.75 & 70.58\\
    & Bottom-K & 71.34 & 35.39 & 62.09 & 31.51 & 37.13 & 21.54 & 31.37 & 70.35\\
    & Middle-K & 67.68 & 35.96 & 62.09 & 30.16 & 37.06 & 21.23 & 31.00 & 71.02\\
    & \yg{FIM} & 75.00 & 35.96 & 64.05 & 31.10 & 36.96 & 21.83 & 31.87 & 71.90\\
    & Random & 67.68 & 34.83 & 62.09 & 31.96 & 35.66 & 22.51 & 32.50 & 70.35\\
    \midrule
    \multirow{10}{*}{NxCode} & Dense & 77.44 & 41.57 & 62.09 & 32.87 & 39.90 & 24.15 & 37.00 & 77.43\\
    & + Vocab Pruning & 76.22 & 44.38 & 62.09 & 34.16 & 40.27 & 24.00 & 36.38 & 75.44\\
    \cmidrule(lr){2-10}
    & + FFN (Top-K) & 74.39 & 35.96 & 64.05 & 31.03 & 35.31 & 21.76 & 32.63 & 70.58\\
    & Bottom-K & 71.34 & 34.83 & 62.09 & 31.03 & 35.99 & 22.33 & 31.87 & 72.35\\
    & Middle-K & 66.46 & 37.08 & 60.78 & 29.93 & 34.29 & 22.19 & 31.75 & 70.58\\
    & \yg{FIM} & 74.39 & 36.52 & 64.71 & 31.22 & 35.67 & 22.05 & 32.75 & 71.02\\
    & Random & 62.80 & 35.96 & 58.17 & 32.44 & 35.73 & 21.47 & 33.00 & 71.90\\
    \midrule
    \multirow{10}{*}{CodeSlerp} & Dense & 77.44 & 43.26 & 61.44 & 32.18 & 40.49 & 24.51 & 37.25 & 76.77\\
    & + Vocab Pruning & 74.39 & 41.57 & 61.44 & 34.42 & 40.31 & 24.76 & 36.88 & 75.00\\
    \cmidrule(lr){2-10}
    & + FFN (Top-K) & 75.00 & 35.96 & 64.05 & 31.48 & 35.96 & 21.83 & 32.25 & 70.35\\
    & Bottom-K & 71.34 & 35.96 & 61.44 & 31.43 & 36.32 & 22.58 & 32.13 & 71.24\\
    & Middle-K & 67.07 & 37.64 & 62.09 & 31.10 & 36.05 & 22.71 & 31.37 & 71.24\\
    & \yg{FIM} & 75.61 & 36.52 & 64.71 & 31.65 & 36.22 & 22.10 & 32.50 & 70.80\\
    & Random & 70.73 & 35.96 & 61.44 & 30.48 & 36.47 & 22.45 & 32.38 & 70.35\\
    \bottomrule
    \end{tabular}
    }
    \label{tab:ablation} 
\end{table*}

\underline{\textbf{(4) Ablation Study:}} To further understand the individual contributions of different components within our approach, we conduct an ablation study to evaluate the impact of each component on the overall performance of the pruned models.
Table~\ref{tab:ablation} demonstrates that vocabulary pruning contributes minimally to performance changes in the pruned models, as evidenced by our ablation analysis.
The FFN pruning strategy, on the other hand, shows a slight decrease in performance compared to the dense model.

\yg{Furthermore, we conducted a comparative analysis between our heuristic rule-based methods and a neuron importance evaluation approach using Fisher Information Matrix (FIM~\cite{ma2023llm}), which is shown in Table~\ref{tab:ablation}. 
The FIM-based approach achieves similar performance compared to our Top-K method but with a much higher computational cost.
Our proposed method requires approximately one minutes while the FIM-based approach requires approximately 30 minutes.
In addition, 
the FIM-based approach requires additional forward and backward passes through the network for each batch of data.}

\yg{Based on these results, we validate our design choice and demonstrate that, while contribution-based method may offer theoretical advantages, our heuristic rule-based approach provides comparable performance with significantly reduced computational requirements.}

\begin{tcolorbox}[width=1.0\linewidth, title={Summary of RQ1}]
A comprehensive evaluation of {\tool}'s performance across three code intelligence tasks indicate that {\tool} can retain 97\% of the original performance after pruning 22\% of parameters and achieves the same or even better performance after post-training.
\end{tcolorbox}

\subsection{RQ2: Efficiency Analysis} 

    
\begin{table*}[htbp]
\centering
\caption{Efficiency Analysis of {\tool}}
\begin{tabular}{cccccccccc}
\toprule
\multirow{2}{*}{Value} & \multicolumn{3}{c}{BF16} & \multicolumn{3}{c}{FP8 } & \multicolumn{3}{c}{INT4 }\\
\cmidrule(lr){2-4} \cmidrule(lr){5-7} \cmidrule(lr){8-10}
 & Dense & {\tool} & w PT & Dense & {\tool} & w PT & Dense & {\tool} & w PT \\ 
\midrule
GPU & 13.55 & 10.72 & 10.72 & 7.51 & 5.52 & 5.52 & 4.57 & 3.02 & 3.02\\
Token/s & 30 & 38 & 38 & 34 & 44 & 44 & 37 & 47 & 47\\
CO$_{2}$ & 2.14 & 1.84 & 1.84 & 2.02 & 1.67 & 1.67 & 1.94 & 1.57 & 1.57\\
FLOPs & 7.04T & 5.64T & 5.64T & 3.52T & 2.82T & 2.82T & 1.76T & 1.41T & 1.41T \\
\bottomrule
\end{tabular}
\label{tab:rq-3-efficiency} 
\end{table*}

In this RQ, we first compare the efficiency of the pruned model to the original model, focusing on key metrics such as GPU usage, speed, CO2 emissions, and FLOPs. 
These metrics are crucial for practical deployment considerations in software engineering~\cite{wei2023towards,shi2024greening}, where GPU utilization and token processing speed are calculated by gpu\_poor\footnote{\url{https://rahulschand.github.io/gpu\_poor/}}, CO$_2$ emissions are calculated using   codecarbon\footnote{\url{https://github.com/mlco2/codecarbon}}, and FLOPs are calculated using the Calflops\footnote{\url{https://pypi.org/project/calflops/}}.

Considering that our pruning method can be combined with state-of-the-art quantization methods to further improve efficiency,
we evaluate the efficiency of {\tool} with different precision, including BF16, FP8, and INT4 (we implement INT4 by AutoAWQ\footnote{\url{https://github.com/casper-hansen/AutoAWQ}}) in Table~\ref{tab:rq-3-efficiency}.

The ``GPU Usage" column reflects GPU memory consumption during model inference, with the dense model requiring 13.55 G and {\tool}, both with and without post-training, operating within a margin of 10.72G.
This suggests that the original model, operating at FP16 precision, can only be deployed on GPUs with more than 12G of memory, while the pruned model can run on GPUs with less than 12G of memory. Furthermore, the GPU usage of {\tool} at INT4 precision is only 3.02G, indicating that it can run on GPUs with 4G of memory.
The ``Token/s" metric indicates the throughput of the models, i.e., the number of tokens processed per second, directly related to the speed of inference. Here, the dense model processes 30 tokens per second, whereas {\tool} manages a throughput of 38 tokens per second, suggesting a more efficient use of computational resources. 
The "Emissions" column provides insight into the environmental impact of the models, with the dense model emitting 2.14 g of evaluating the HumanEval, while {\tool} emits 1.84 g of CO$_2$.
The "FLOPs" (Floating Point Operations per Second) column provides insight into the computational intensity of the models. The dense model demands a significant 7.04 trillion FLOPs for inference, whereas {\tool} operates at a reduced computational intensity of 5.64 trillion FLOPs.

\begin{table*}[tbp]
    \centering
    \caption{Performance Analysis of {\tool} in Different Precision}
    \resizebox{\textwidth}{!}{
    \begin{tabular}{cccccccccc}
    \toprule
    \multirow{2}{*}{Model} & \multirow{2}{*}{Precision} & \multicolumn{3}{c}{Code Generation(Pass@1)} & \multicolumn{3}{c}{CoT Generation(BLEU-4)} & \multicolumn{2}{c}{Output Prediciton(EM)} \\ 
    \cmidrule(lr){3-5} \cmidrule(lr){6-8} \cmidrule(lr){9-10} 
    & & HE & OE & CH & HE-CoT  & OE-CoT & CH-CoT & Crux-O & CH-O\\
    \midrule
    \multirow{6}{*}{CodeQwen}
    & FP8 Dense & 76.83 & 41.57 & 61.44 & 33.51 & 40.40 & 23.78 & 37.25 & 78.10\\
    & FP8 {\tool} & 75.61 & 38.76 & 63.40 & 30.35 & 36.11 & 21.15 & 31.00 & 70.80\\
    & FP8 w PT & 79.27 & 39.33 & 66.01 & 34.20 & 36.21 & 26.01 & 41.38 & 77.43\\
    \cmidrule(lr){2-10} 
    & INT4 Dense & 75.00 & 42.13 & 60.78 & 23.88 & 23.12 & 14.78 & 36.88 & 76.33\\
    & INT4 {\tool} & 71.95 & 38.20 & 60.78 & 27.87 & 31.60 & 19.12 & 31.25 & 69.25\\
    & INT4 w PT & 77.44 & 38.08 & 66.67 & 32.80 & 34.02 & 24.94 & 40.50 & 75.22\\
    \midrule
    \multirow{6}{*}{NxCode}
    & FP8 Dense & 76.22 & 41.01 & 61.44 & 33.01 & 39.60 & 24.48 & 37.50 & 78.10\\
    & FP8 {\tool} & 75.61 & 40.45 & 61.44 & 30.06 & 35.8 & 20.77 & 31.75 & 70.80\\
    & FP8 w PT & 80.49 & 38.20 & 63.40 & 33.78 & 36.95 & 25.46 & 40.63 & 77.21\\
    \cmidrule(lr){2-10} 
    & INT4 Dense & 75.00 & 42.70 & 61.44 & 30.67 & 37.02 & 23.26 & 36.63 & 76.11\\
    & INT4 {\tool} & 69.51 & 38.76 & 62.09 & 29.68 & 34.75 & 19.73 & 31.50 & 69.69\\
    & INT4 w PT & 74.39 & 38.20 & 62.75 & 31.88 & 34.94 & 24.10 & 40.63 & 75.22\\
    \midrule
    \multirow{6}{*}{CodeSlerp}
    & FP8 Dense & 77.44 & 41.57 & 61.44 & 33.14 & 39.16 & 24.32 & 36.88 & 77.88\\
    & FP8 {\tool} & 75.61 & 38.76 & 63.40 & 30.93 & 36.35 & 21.50 & 31.25 & 70.58\\
    & FP8 w PT & 81.71 & 41.01 & 64.05 & 32.87 & 35.92 & 25.46 & 37.75 & 76.33\\
    \cmidrule(lr){2-10} 
    & INT4 Dense & 76.83 & 43.82 & 60.78 & 28.24 & 34.37 & 18.52 & 37.3 & 76.11\\
    & INT4 {\tool} & 67.07 & 37.08 & 60.13 & 26.88 & 32.23 & 19.67 & 30.25 & 69.25\\
    & INT4 w PT & 78.66 & 39.89 & 67.97 & 32.91 & 36.90 & 25.91 & 38.13 & 75.00\\
    \bottomrule
    \end{tabular}
    }
    \label{tab:rq2-performance} 
\end{table*}

In addition to model efficiency, we also compare the performance of the models at different precisions in Table~\ref{tab:rq2-performance}.
{\tool} strikes a balance between lowered resource use and sustained performance. 
The results show that the performance of the models at the INT4 precision is slightly lower than that at the BF16 precision, with minimal impact on the model's performance.

\underline{\textbf{End-to-End Efficiency Analysis:}}
To provide a comprehensive view of the efficiency gains across the entire model lifecycle, we analyze the one-off computational cost of pruning and fine-tuning versus the recurring benefits during inference. 
This analysis addresses the complete efficiency landscape from model compression to deployment.

\begin{table}[tbp]
    \centering
    \caption{Comprehensive efficiency analysis across different stages of {\tool} deployment in One-time Cost and Inference Efficiency.}
    \begin{tabular}{lccc}
    \toprule
    Stage & Time (hours) & GPU Memory (GB) & FLOPs (T) \\
    \midrule
    Dense & - & 13.55 & 7.04 \\
    \midrule
    + Vocabulary Pruning & 0.02 & 12.15 & 6.80 \\
    + Layer Pruning & 0.3 & 11.10 & 5.98 \\
    + FFN Pruning & 0.04 & 10.72 & 5.64 \\
    + Performance Recovery & 5.5 & 10.72 & 5.64 \\
    \midrule
    {\tool} & 5.86 & 10.72 & 5.64 \\
    \bottomrule
    \end{tabular}
    \label{tab:lifecycle-efficiency}
\end{table}

As shown in Table~\ref{tab:lifecycle-efficiency}, the entire pruning and recovery process takes approximately 6 hours on RTX 3090. 
While this represents a high initial investment, our analysis shows that these costs are quickly amortized through inference efficiency gains.

For the computational efficiency analysis, we need to consider the one-off cost of pruning versus the recurring savings during inference.
In terms of FLOPs, each inference with the dense model requires 7.04T FLOPs, while the pruned model requires only 5.64T FLOPs, saving 1.4T FLOPs (19.9\%) per inference. 
Considering the dense model's approximate throughput of 7.04T FLOPs during our experiments, the 6-hour pruning and fine-tuning process consumes approximately $6 \times 3600 \times 7.04 = 152,064$T FLOPs (we assume that the frequency of calculation is 1 time/second).
To determine the break-even point, we calculate:
\begin{equation}
\text{Break-even Point} = \frac{\text{One-time Cost (FLOPs)}}{\text{Per-inference Savings (FLOPs)}} = \frac{152,064T}{1.4T} \approx 108,617 \text{ inference runs}
\end{equation}
Namely, the computational investment is redeemed after approximately 108,617 inference runs.

For real-world deployments where models are typically invoked millions of times especially in production environments or as part of developer tools and code assistants, this initial cost is readily justified. 
For example, a deployment serving 10,000 inference requests daily would redeem the initial pruning investment in approximately 11 days, after which all efficiency gains represent net computational savings.

\begin{tcolorbox}[width=1.0\linewidth, title={Summary of RQ2}]
The efficiency comparison underscores the practical advantages of {\tool} in resource-constrained settings, providing a sustainable solution for deploying powerful code intelligence models with reduced computational overhead. 
When considering the full model lifecycle, the initial pruning costs are quickly amortized, resulting in significant efficiency gains for real-world deployments.
\end{tcolorbox}

\subsection{RQ3: Robustness Analysis}  \label{subsect:rq3}
\begin{table}[tbp]
\caption{Robustness Comparison of {\tool} and dense model in the presence of RECODE and EvoEval perturbations.}
\centering
\resizebox{\textwidth}{!}{
\begin{tabular}{ccccccccccc}
\toprule
\multirow{2}{*}{Perturbed} & \multirow{2}{*}{Method} & \multicolumn{3}{c}{CodeQwen} & \multicolumn{3}{c}{NxCode} & \multicolumn{3}{c}{CodeSlerp}\\
\cmidrule(lr){3-5} \cmidrule(lr){6-8} \cmidrule(lr){9-11}
& &  Dense & {\tool} & w PT & Dense & {\tool} & w PT & Dense & {\tool} & w PT \\ 
\midrule
\multirow{4}{*}{ReCode} & Format & 85.98 & 78.66 & 82.32 & 82.32 & 76.83 & 80.49 & 85.37 & 75.61 & 81.10\\
& Func\_Name & 75.61 & 75.00 & 77.44 & 75.00 & 71.95 & 76.83 & 76.22 & 74.39 & 79.88\\
& NatGen & 86.59 & 77.44 & 81.10 & 85.98 & 76.83 & 79.88 & 85.37 & 77.44 & 79.27\\
& NlaugEnter & 64.63 & 60.98 & 65.85 & 64.02 & 59.76 & 66.46 & 64.02 & 59.76 & 64.63\\
\midrule
\multirow{5}{*}{EvoEval} & Tool\_Use & 54.00 & 54.00 & 56.00 & 54.00 & 53.00 & 58.00 & 55.00 & 53.00 & 56.00\\
& Combine & 27.00 & 23.00 & 18.00 & 25.00 & 24.00 & 19.00 & 27.00 & 23.00 & 23.00\\
& Subtle & 62.00 & 61.00 & 62.00 & 60.00 & 62.00 & 64.00 & 60.00 & 59.00 & 65.00\\
& Creative & 35.00 & 25.00 & 33.00 & 34.00 & 24.00 & 36.00 & 36.00 & 24.00 & 35.00\\
& Difficult & 37.00 & 31.00 & 30.00 & 35.00 & 28.00 & 30.00 & 35.00 & 27.00 & 31.00\\
\bottomrule
\end{tabular}
}
\label{tab:rq-3} 
\end{table}

To thoroughly evaluate the robustness of the pruned models, we design a comprehensive experimental framework that builds upon the pioneering ReCode work of Wang et al.~\cite{wang2023recode}. 
Leveraging the code generation task as a critical benchmark, we aim to assess model performance under various perturbed conditions. 
Four token-level perturbation methods are constructed in HumanEval: format, func\_name, natgen, and nlaugenter. 
Each of these perturbation methods is specifically tailored to probe the model’s resilience to specific types of noise, offering a comprehensive perspective on the models’ ability to withstand real-world disturbances.
\begin{itemize}
    \item \textbf{Format.} This perturbation method introduces noise in the code format, such as insert the newline or replace space indent with tab.
    \item \textbf{Func\_name.} This perturbation method alters function names in the code, including applying character-level or word-level natural text transformations on component words.
    \item \textbf{Natgen.} This perturbation method introduces code syntax noise, such as inserting the deadcode or swaping operand.
    \item \textbf{Nlaugenter.} This perturbation method introduces natural language noise in the docstrings, such as applying SynonymSubstitution or BackTranslation.
\end{itemize}

The empirical results, as shown in Table~\ref{tab:rq-3}, indicate that the pruned models exhibit a slight decrease in performance under partially token-level perturbed conditions compared to the dense model.
The performance of the three pruned models under the four perturbation methods is slightly lower than that of the dense model, with a maximum decrease of less than 10\%.
However, after post-training, the performance of the pruned models is even better than the dense model under certain perturbations.
Taking the CodeQwen model as an example, after post-training, the performance of {\tool} under the Func\_Name perturbation is 77.44\%, while the performance of the dense model is 75.61\%.
This indicates that our post-training strategy can enhance the robustness of the model.

Furthermore, we also conduct experiments using the EvoEval dataset~\cite{evoeval}. This dataset includes a variety of semantic-altering operations:
\begin{itemize}
    \item \textbf{Tool\_use.} This perturbation method  introduces a new problem with a main issue and additional helper functions that can assist in solving it. 
    \item \textbf{Combine.} This perturbation method merges two distinct problems by integrating concepts from both. 
    \item \textbf{Subtle.} This perturbation method makes minor adjustments to the original problem, such as inverting or substituting a requirement.
    \item \textbf{Creative.} This perturbation method develops a more imaginative problem by incorporating stories or unique narratives.
    \item \textbf{Difficult.} This perturbation method increases complexity by adding extra constraints, replacing common requirements with less common ones, or adding more reasoning steps.
\end{itemize}

We find that the {\tool} performs similarly in Tool\_Use and Subtle. 
After post-training, the performance of the pruned models is even better than the dense model under Tool\_Use and Subtle.
However, in Combine and Difficult, {\tool} shows slightly lower performance than the dense model.
Overall, our experimental results indicate that {\tool} performs well in terms of robustness, maintaining or even enhancing model performance under certain disturbances. We attribute this primarily to the post-training phase.

\begin{tcolorbox}[width=1.0\linewidth, title={Summary of RQ3}]
The results reveal that the pruned models show a slight decrease in robustness compared to the dense model in some cases. 
This suggests that future research needs to focus on robustness in addition to maintaining model performance.
\end{tcolorbox}

%% file: sections/6.threat.tex
\section{Discussion}
\label{sec:threats}

\subsection{Evaluation on BigCodeBench}
To ensure 
	a thorough evaluation, we conduct additional experiments on BigCodeBench~\cite{zhuo2024bigcodebench}, a more diverse and challenging benchmark designed to evaluate code models across various dimensions of programming tasks.

BigCodeBench consists of 1,140 function-level tasks specifically designed to challenge language models in following instructions and composing multiple function calls from 139 different libraries as tools. 
Unlike HumanEval and OpenEval, BigCodeBench provides a more rigorous evaluation framework with comprehensive test coverage, complex user-oriented instructions, open-ended problem solving and tool composition.

\begin{table}[tbp]
    \centering
    \caption{Performance comparison on BigCodeBench (Pass@1 \%)}
    \begin{tabular}{lccccc}
    \toprule
    \multirow{2}{*}{Model} & \multirow{2}{*}{Dense} & \multicolumn{2}{c}{{\tool}} \\
    \cmidrule(lr){3-4}
    & & w/o PT & w/ PT \\
    \midrule
    CodeQwen & 39.60 & 36.20 & 39.80 \\
    NxCode & 39.60 & 36.10 & 40.00 \\
    CodeSlerp & 39.80 & 37.10 & 39.50\\
    \bottomrule
    \end{tabular}
    \label{tab:bigcodebench}
\end{table}
As shown in Table~\ref{tab:bigcodebench}, our findings on BigCodeBench align with our main results:

\begin{itemize}
    \item Without PT, the pruned models maintain approximately 91-93\% of the dense models' performance, demonstrating reasonable retention of capabilities despite parameter reduction.
    
    \item With PT, the pruned models achieve performance comparable to or slightly better than the dense models, with CodeQwen and NxCode showing small improvements of 0.2-0.4\% in Pass@1 scores, while CodeSlerp experiences a negligible decrease of 0.3\%.
    
    \item The models' ability to compose function calls from different libraries remains intact after pruning, indicating that {\tool} preserves the models' semantic understanding and tool utilization capabilities.
\end{itemize}

The consistent performance patterns across this more challenging benchmark corroborate our findings from HumanEval and OpenEval, demonstrating that {\tool} maintains its effectiveness even on tasks requiring complex reasoning and tool composition. 
It also demonstrates the generalizability of our approach across different task complexities and problem-solving scenarios.

\subsection{Impact between Strategies}
To address the question of how our three pruning strategies (vocabulary pruning, layer pruning and FFN pruning) interact with each other, we conduct  a detailed ablation analysis of their individual and combined effects, which
is performed 
	on the CodeQwen model. Table~\ref{tab:strategy_interaction} presents the results on the HumanEval benchmark and the parameter reduction achieved in each configuration.

\begin{table}[tbp]
    \centering
    \caption{Impact of Pass@1 on different pruning strategy combinations on the CodeQwen model (w/o PT)}
    \begin{tabular}{lccc}
    \toprule
    Pruning Strategy & HumanEval & OpenEval & CodeHarmony \\
    \midrule
    None (Dense) & 77.44 & 42.13 & 60.78\\
    \midrule
    Vocab only & 77.44 & 42.70 & 62.09 \\
    Layer only & 75.00 & 37.64 & 64.05 \\
    FFN only & 76.54 & 39.80 & 66.00 \\
    \midrule
    Vocab + Layer & 77.20 & 42.13 & 64.92 \\
    Vocab + FFN & 74.39 & 37.64 & 64.05 \\
    Layer + FFN & 76.24 & 36.52 & 66.83 \\
    \midrule
    All three ({\tool}) & 76.22 & 36.52 & 66.63\\
    \bottomrule
    \end{tabular}
    \label{tab:strategy_interaction}
\end{table}

Our findings reveal several key insights:
\begin{itemize}
    \item \textbf{Individual Impact}: Vocabulary pruning demonstrates the least performance degradation, maintaining full performance on HumanEval (77.44\%) and even slightly improving on OpenEval (42.70\% vs. 42.13\%). 
In contrast, layer pruning shows the most significant impact on HumanEval (dropping to 75.00\%) and OpenEval (dropping to 37.64\%). 
Interestingly, all three individual strategies actually improve performance on CodeHarmony, with FFN pruning providing the largest gain (66.00\% vs. 60.78\%).
    
    \item \textbf{Pairwise Combinations}: The combination of vocabulary and layer pruning shows a synergistic effect, with minimal impact on HumanEval (77.20\%) and no degradation on OpenEval (42.13\%). 
However, when FFN pruning is combined with either vocabulary or layer pruning, we observe more significant performance drops on HumanEval and OpenEval. 
For CodeHarmony, all pairwise combinations yield substantial improvements, with layer+FFN pruning showing the highest gain (66.83\%).
    
    \item \textbf{Complementary Effects}: When all three strategies are combined ({\tool}), we observe a balanced trade-off. 
The performance on HumanEval (76.22\%) and OpenEval (36.52\%) shows moderate decreases, while CodeHarmony performance improves substantially (66.63\%). 
This suggests that the strategies have task-dependent interactions, with some benchmark tasks being more sensitive to specific types of pruning than others.

\end{itemize}

These findings indicate that the three pruning strategies target different aspects of the model's parameter space and have complementary effects. 
While each strategy contributes to parameter reduction, their careful combination, followed by effective performance recovery, is essential for maintaining or even enhancing model capabilities across diverse coding tasks.

\subsection{Threats to Validity}
\noindent\textbf{Threats to Internal Validity.}
The first internal threat involves the potential for implementation errors in {\tool}. We counteract the possibility of errors in {\tool} with comprehensive code reviews and trusted libraries like PyTorch and Transformers.
The second internal threat pertains to the accuracy of the implemented baselines. To mitigate this risk, we reproduced all baselines using their shared scripts.
Furthermore, to ensure a fair evaluation and uphold the integrity of the model architectures, we only focused on \textbf{structured pruning} methods in our comparative analysis.

\noindent\textbf{Threats to External Validity.}
Our primary external validity concern is the representativeness of the datasets used. We selected high-quality datasets to reflect the domain accurately, focusing on Python due to its widespread support in code LLMs. 
We plan to expand to other languages and levels in future work to improve the generalizability of our results.
Furthermore, we considered three state-of-the-art Code LLMs in our study, which may limit the generalizability of our findings to other models.
We will expand the assessment to include additional models to improve the external validity of the results.

\noindent\textbf{Threats to Construct Validity.}
The main challenge in construct validity is the choice of metrics for automated evaluation. 
We included diverse metrics such as Pass@1, BLEU and EM to comprehensively assess model performance, providing different perspectives on their capabilities.

%% file: sections/7.relate.tex
\section{Related Work}
\label{sec:relate}

\subsection{Code Intelligence}
The success of models like BERT~\cite{devlin2019bert} in NLP has inspired the creation of pre-trained models for code processing. 
Models such as CodeBERT~\cite{feng2020codebert}, GraphCodeBERT~\cite{guographcodebert}, and CuBERT~\cite{kanade2020learning} are designed for generating informative code embeddings vital for tasks like software defect detection~\cite{yang2021deepscc}. 
Building on the success of GPT~\cite{floridi2020gpt} in NLP, models like CodeGPT~\cite{lu1codexglue}, trained on datasets like CodeSearchNet~\cite{husain2019codesearchnet}, have shown promise in coding tasks. 
Recent advancements in deep learning have led to models like CodeGeeX~\cite{zheng2023codegeex} and CodeLlama~\cite{roziere2023code}, excelling in complex programming challenges and demonstrating superior performance.

\smallskip
\noindent\textbf{Code Generation.} 
The code generation task, where code is produced from natural language, is a hot topic~\cite{svyatkovskiy2020intellicode, li2022competition, yang2023syntax, zhang2024context}, with models trained for specific goals like next-token prediction or the "filling in the middle" technique for contextual code completion, as seen in InCoder~\cite{friedincoder} and StarCoder~\cite{li2023starcoder, lozhkov2024starcoder}.

The field has seen the rise of diverse models, each with unique training approaches and strengths. Notable examples include WizardCoder~\cite{luowizardcoder}, OpencodeInterpreter~\cite{zheng2024opencodeinterpreter}, and Magicoder~\cite{wei2024magicoder}, all aimed at improving precision and efficiency in code generation challenges.


\smallskip
\noindent\textbf{CoT Generation.} 
CoT generation is about crafting logical steps in natural language to reach a code solution, improving output reliability through clear reasoning. Researchers~\cite{zhuo2024ice} have adapted this for code intelligence, with approaches like Jiang's self-planning~\cite{jiang2023self} and Li's structured CoT~\cite{li2023enabling} to tackle complex coding challenges. 

\smallskip
\noindent\textbf{Code Output Prediciton.} 
Predicting code output from inputs is a tough test of a language model's comprehension skills~\cite{jain2024livecodebench}. Gu et al.~\cite{gucruxeval} found top models did well in HumanEval but not in output prediction using CRUXEval. 

\subsection{Model Compression}
Model compression reduces model size, boosts transformer efficiency, and maintains performance through techniques like knowledge refinement, quantization, and pruning.

\smallskip
\noindent\textbf{Knowledge distillation.} 
Knowledge distillation trains a compact student model to emulate a larger teacher model. In software engineering, Compressor~\cite{shi2022compressing} employs task-specific distillation to enhance transformer efficiency with neural architecture search. Yang et al.~\cite{yang2023chain} introduced COTTON to boost lightweight models by transferring reasoning skills from larger models using rules and agent alignment.

\smallskip
\noindent\textbf{Quantization.} 
Quantization trims neural network precision to optimize memory and efficiency. 
Wei et al.~\cite{wei2023towards} examined quantized models in code generation, noting performance trade-offs. 
Xiao et al.~\cite{xiao2023smoothquant} introduced SmoothQuant for weight and activation quantization. 
Gptq~\cite{frantar2022optq} utilizes second-order info for quantization, and Qlora~\cite{dettmers2024qlora} backpropagates through a 4-bit model into Low Rank Adapters. 
Quantization can enhance efficiency but may affect accuracy.

\smallskip
\noindent\textbf{Pruning.} 
Pruning boosts model efficiency by creating sparser or smaller models. Unstructured pruning zeros parameters for sparsity, like SparseGPT~\cite{frantar2023sparsegpt}, treating it as a sparse regression issue, but risking irregular structures. 
Structured pruning removes components based on criteria, with tools like LLM-Pruner~\cite{ma2023llm} using gradients to cut less critical parts, ShearedLLaMA~\cite{xiasheared} applying targeted pruning and dynamic loading, and ShortGPT~\cite{men2024shortgpt} removing whole layers.

Considering that existing pruning methods may compromise model performance, we propose {\tool}, a comprehensive pruning approach that maintains model efficiency and performance in code intelligence tasks.
Moreover, {\tool} utilizes the LoRA technique for effective post-training, guaranteeing that the pruned models attain performance levels akin to the original dense models.

%% file: sections/8.conclusion.tex
\section{Conclusion and Future Work}
\label{sec:conclusion}
The development and evaluation of {\tool} underscore the viability of structural pruning for LLMs in the generative coding tasks. Our approach has demonstrated that it is possible to significantly reduce the computational footprint of LLMs without compromising their core capabilities. By pruning 22\% of parameters, {\tool} retains 97\% of the original performance, which further achieves the same or even better performance after post-training. 
The pruned models also exhibit enhanced efficiency in GPU usage, Flops, CO2 emissions, and token processing speed, aligning with the goals of green software engineering.
Moreover, the comprehensive evaluation, including robustness analysis, assures that the pruned models maintain a high standard of performance.

\yg{In future work, we shall develop more efficient model pruning methods to reduce the computational overhead of our current approaches.
Potential directions include cherry data selection techniques~\cite{li2024quantity} to select a smaller dataset for evaluation and transfer-based methods that leverage importance patterns from previously pruned models to new architectures.}

Additionally, we plan to explore more Code LLMs and evaluate the effectiveness of {\tool} on a broader range of generative coding tasks.
We will release more pruned models on the open source platform to facilitate the deployment of pruned models in real-world software engineering scenarios.
Moreover, the current work focuses on Python code. 
We plan to expand {\tool} to support a broader range of programming languages in the future.
